\documentstyle[12pt,epsfig]{article}
\begin{document}
\def\e{\epsilon}
\def\d{{\rm d}}
\def\Li{{\rm Li}}
\def\zcut{z_{\rm cut}}
\def\ycut{y_{\rm cut}}
\def\ymin{y_{\rm min}}
\def\scut{y_{\rm cut}}
\def\smin{s_{\rm min}}
\def\yqqb{y_{q\bar{q}}}
\def\yqg{y_{q\gamma}}
\def\yqbg{y_{\bar{q}\gamma}}
\def\Dg{D_{g\to \gamma}}
\def\DBqp{D^B_{q\to \gamma}}
\def\DBq{D^B_{q\to \gamma}}
\def\DBqpp{D^B_{q'\to \gamma}}
\def\DBgp{D^B_{g\to \gamma}}
\def\Dq{D_{q\to \gamma}}
\def\Di{D_{i\to \gamma}}
\def\Dqp{D_{q'\to \gamma}}
\def\DBa{D^B_{a\to \gamma}}
\def\Pqqzero{P^{(0)}_{q\to q}}
\def\Pqgzero{P^{(0)}_{q\to g}}
\def\Pgqzero{P^{(0)}_{g\to q}}
\def\Pqpzero{P^{(0)}_{q\to \gamma}}
\def\Pqqone{P^{(1)}_{q\to q}}
\def\Pqpone{P^{(1)}_{q\to \gamma}}
\def\Pgpone{P^{(1)}_{g\to \gamma}}
\def\Pqg{P_{q\to g}}
\def\Pqp{P_{q\to \gamma}}
\def\Gqp{\Gamma_{q\to \gamma}}
\def\Gqpp{\Gamma_{q'\to \gamma}}
\def\Ggp{\Gamma_{g\to \gamma}}
\def\Gqg{\Gamma_{q\to g}}
\def\Ggq{\Gamma_{g\to q}}
\def\Gij{\Gamma_{i\to j}}
\def\Gqq{\Gamma_{q\to q}}
\def\Ggg{\Gamma_{g\to g}}
\def\Gqpq{\Gamma_{q'\to q}}
\def\Gqqp{\Gamma_{q\to q'}}
\def\aqp{a_{q\to \gamma}}
\def\agp{a_{g\to \gamma}}
\def\calDq{{\cal D}_{q\to \gamma}}
\def\calDg{{\cal D}_{g\to \gamma}}
\def\aqed{\left(\frac{\alpha e_q^2}{2\pi}\right)}
\def\as{\left(\frac{\alpha_s}{2\pi}\right)}
\def\4pi2{\left(\frac{4\pi\mu^2}{M^2}\right)^{2 \e}}
\def\coup{\frac{1}{\Gamma(1-\e)^2}\4pi2\as\aqed}

\pagestyle{myheadings}

\begin{titlepage}
\vspace*{-1cm}
\begin{flushright}
DTP/98/26   \\
DESY/98/68 \\
June 1998 \\
\end{flushright}                                
\vskip 1.cm
\begin{center}                                                             
{\Large\bf
Final state photon production at LEP }
\vskip 1.3cm
{\large A.~Gehrmann--De Ridder$^a$ and  E.~W.~N.~Glover$^b$}
\vskip .2cm
$^a$~{\it DESY, Theory Group, D-22603 Hamburg, Germany} \\
$^b$~{\it Department of Physics, University of Durham,
Durham DH1 3LE, England }
\vskip 2.3cm   
\end{center}      
\begin{abstract}
We present a detailed study of photon production
in hadronic events in electron-positron 
annihilation at LEP energies.
We show that estimates of the inclusive photon spectrum using
the quark-to-photon fragmentation function 
determined using the
ALEPH `photon' +~1 jet data agree well with the observations of the OPAL collaboration.
This agreement shows that the photon fragmentation function 
determined in this way can be used for inclusive observables.
We also compare next-to-leading order and beyond leading
logarithm predictions obtained using the numerically resummed
solutions of the fragmentation function evolution equation of
Bourhis, Fontannaz and Guillet and Gl\"uck, Reya and Vogt with the data.
Moreover, in order to check the general behaviour of the 
fragmentation function, we consider an analytic series expansion 
in the strong coupling. 
We find that the parameterizations are inaccurate at large $x$ values.
While the OPAL data is in broad agreement with estimates
based on any of these approaches,
the ALEPH data prefers the resummed BFG parameterization.
Finally, there is some ambiguity as to whether the fragmentation function 
is treated as ${\cal O}(\alpha)$ or ${\cal O}(\alpha/\alpha_s)$.
We show that at present this ambiguity affects mainly the prediction
for the `photon' +~1 jet rate at large $z$.
\end{abstract}                                                                
\vfill
\end{titlepage}
\newpage                                                                     

\section{Introduction} 
\setcounter{equation}{0}
\label{sec:intro}
The production of hard photons in hadronic processes 
provides an important testing ground for QCD. 
For example, direct photon production in $p\bar{p}$ collisions is used     
to extract information on the gluon content of the proton,
while the presence of photons in the final state 
represents an important background source in many searches 
for new physics. A good understanding
of direct photon production within the context of 
the Standard model is therefore essential.
  
Photons produced in hadronic collisions 
can have two possible origins: the {\it direct} radiation of a 
photon off a primary quark and the {\it fragmentation} of a 
parton $i$ into a photon carrying a sizeable fraction $x$, of the parton  
energy. While the former direct process constitutes a short-distance effect 
and can be calculated within the framework of the Standard Model, 
the latter is primarily a long distance process. It is described by 
the process-independent parton-to-photon fragmentation function 
$D_{i \to \gamma}(x,\mu_{F})$, which cannot be calculated 
using perturbative methods but which 
must be determined from experimental data. 
The evolution of $D_{i \to \gamma}(x,\mu_{F})$ with the factorization scale 
$\mu_{F}$ can however be determined by perturbative QCD.

The most promising environment for a determination of the quark-to-photon 
fragmentation function $\Dq(x,\mu_{F})$ is the study of 
photon production in electron-positron annihilation into hadrons. 
Such measurements have been recently presented by the ALEPH \cite{aleph} 
and OPAL \cite{opal} collaborations at CERN.  
Both measurements differ not only in the experimental observable
studied to determine the quark-to-photon fragmentation function, 
but also in the theoretical framework used 
to match direct and fragmentation contributions onto each other.
This makes a comparison of both experiments difficult, 
and has also given rise to the speculation that the ALEPH and OPAL 
analyzes do not probe the same quantity \cite{opal,bfg}.   

The analysis performed by ALEPH is based on the study of 
two jet events in which one of the jets contains a highly  
energetic photon. Correspondingly, 
the fraction of the jet momentum carried by the 
photon within the `photon' jet, $z$, 
is greater than $0.7$. 
These events were defined by the application 
of the Durham jet clustering algorithm \cite{durham} to both
hadronic and electromagnetic clusters. 
A comparison between the observed rate and the $\cal
{\cal O}(\alpha)$ \cite{aleph,andrew} or ${\cal O}(\alpha \alpha_{s})$ 
\cite{letter} theoretical estimates
yielded a first determination of 
the quark-to-photon fragmentation function accurate 
at leading and next-to-leading order.
The theoretical basis on which the measurement of the `photon' +~1 jet rate 
relies, is an explicit counting of powers of the strong coupling $\alpha_{s}$ 
in both the direct and the fragmentation contributions,
and where no resummation of $\ln \mu_{F}^2$ is performed.
We shall refer to this theoretical framework as the fixed order approach.

More recently, the OPAL collaboration has measured the inclusive 
photon distribution for final state photons 
with energies as small as 10 GeV.
This corresponds to the photon carrying a fraction of the
beam momentum, $x_{\gamma}$, to be as low as 0.2. 
They have compared their results with the model-dependent predictions 
of \cite{bfg,grv} and found a reasonable agreement in both cases.  
These model predictions are based on a resummation of 
the logarithms of the factorization scale $\mu_{F}$ 
and naturally associate an inverse power of $\alpha_{s}$ 
with all fragmentation contributions. This resummation procedure 
is the conventional approach. 

Although, the quark-to-photon fragmentation function 
is a purely non perturbative object,
its evolution with the factorization scale can be
described within perturbative QCD.
In the fixed order approach, 
a byproduct of the perturbative calculation of the photon 
production cross section  
is an evolution equation which is accurate at a given order. 
This has been presented in some detail in \cite{letter},
and we recall the main features in Section~\ref{sec:ourinclusive}.
In the conventional approach adopted by \cite{bfg,grv} though, 
the quark-to-photon fragmentation function satisfies 
the well-known DGLAP \cite{AP} all order evolution equation and, as    
in the fixed order approach, this solution 
has a non-perturbative ingredient which can only be measured. 
We discuss the main steps in the formal derivation of this solution 
in the conventional formalism in Section \ref{sec:conventional}. 

A priori, it may indeed not be entirely clear that the fragmentation 
function extracted from a measurement of the `photon' +~1 jet rate  
and the fragmentation function arising in the inclusive cross section 
are the same. In fact, the variables used in the definitions of these rates, 
$z$ for the photon rate and $x_{\gamma}$ for the inclusive rate
are generally not equal to each other.
As explained in Section \ref{sec:ourinclusive}, these 
variables coincide however in the purely `collinear' region 
where the genuine non-perturbative effects arise so 
we can clarify this issue and affirm that these two 
fragmentation functions are indeed equal to each other.
Information gained about the quark-to-photon fragmentation function
from the analysis of one observable
can (and ought to) be used to predict other photon cross sections.

So far the ALEPH collaboration have compared their data
with the theoretical predictions obtained in a fixed order formalism  
while the OPAL collaboration have concentrated on 
the theoretical predictions from the conventional framework. 
However, because
the fragmentation functions are universal, it is possible 
to evaluate the `photon' +~1 jet rate and the inclusive rate obtained in 
{\em either} theoretical framework and to compare these predictions with
{\em either} the ALEPH or the OPAL data.
To perform such a comparison, and to see if the data prefers one approach, 
is the purpose of this paper. 

More precisely the paper is organized as follows.
In Section~\ref{sec:ourinclusive}, we discuss 
how the quark-to-photon fragmentation function and the 
one-photon production cross section are determined  
in a fixed order approach and present the results obtained 
for the inclusive rate in this approach. 
In Section~\ref{sec:conventional}, we describe how the quark-to-photon 
fragmentation function is determined in a conventional approach.
In particular we describe how the leading-logarithmic (LL) 
and beyond-leading logarithmic (BLL) solutions are determined 
and compare their numerical parameterizations 
with analytically expanded expressions 
of the quark-to-photon fragmentation function.
Section~\ref{sec:comparison} contains a detailed presentation 
of four possible different approaches to evaluate the one-photon 
production cross sections, one of these being closely related 
to the fixed order approach described in Section~\ref{sec:ourinclusive}. 
The possible definitions of the non-perturbative input 
in either the schemes adopted by Gl\"{u}ck, Reya and Vogt (GRV) in \cite{grv}
or by  Bourhis, Fontannaz and Guillet (BFG) in \cite{bfg}, 
are also discussed in this context.
Moreover we study
the behaviour of the 
parameterizations given in \cite{bfg,grv} in the large $x$ region.  
In Section~\ref{sec:results}, 
we present results obtained following any of 
these four approaches and using either the GRV or BFG schemes 
for the inclusive and `photon' +~1 jet 
cross sections 
and compare them with the OPAL and ALEPH data.    
Finally, Section~\ref{sec:conclusions} contains our conclusions.

\section{Inclusive photon production in the fixed order 
approach}
\setcounter{equation}{0}
\label{sec:ourinclusive}
Let us first consider the general structure of the single photon 
production cross section, fully differential in all quantities, 
\begin{equation}
{\rm d}\sigma
={\rm d}\hat \sigma_\gamma \;+\;  
\sum_{a}{\rm d}\hat \sigma_a \otimes D^{B}_{a\to \gamma}.
\label{eq:gstruc1}
\end{equation}
In this equation which is valid for any single photon production cross section 
there are two contributions. First `prompt' photon production 
arises when the photon is produced directly in the hard interaction. 
Second, the longer distance
fragmentation process occurs when one of the partons produced in the hard
interaction fragments 
into a photon and transfers a fraction of the parent parton 
momentum to the photon.
Each type of parton, $a$, contributes according to the process independent 
parton-to-photon fragmentation functions $D_{a \to \gamma}$ 
and the sum runs over all partons.
At the order we are interested in, the gluon fragmentation functions will
generally be small and can be neglected, as explained below.
The sum therefore runs over all active quark and antiquark flavours.
Moreover,  by charge conjugation, quark and antiquark  fragmentation 
functions are equal. 

The individual terms in eq.~(\ref{eq:gstruc1}) may be divergent and are 
denoted by hatted quantities.
However, through the introduction of a factorization scale $\mu_{F}$,
these terms can be reorganized and the physical cross section can be written  
in terms of finite (but factorization scale dependent) quantities, 
\begin{equation}
{\rm d}\sigma
=
{\rm d}\sigma_\gamma(\mu_F)\;+\;  
\sum_{a}{\rm d}\sigma_a \otimes D_{a\to \gamma}(\mu_F).
\label{eq:gstruc2}
\end{equation}

These two process specific contributions will be defined differently 
for the `photon' + 1~jet and inclusive photon cross sections.
Furthermore, as we shall see in Section \ref{sec:comparison} 
these will be defined differently depending which approach 
is used to evaluate these photon production cross sections. 

We are primarily interested in the expression of the 
inclusive photon production   
evaluated at fixed order up to ${\cal O}(\alpha \alpha_{s})$.
In this context, the hard cross sections 
as well as the non-perturbative fragmentation functions have to 
be considered at most up to ${\cal O}(\alpha \alpha_{s})$.
Although the fragmentation functions are non-perturbative, 
we can nominally assign a power of coupling constants, 
based on counting the couplings necessary to radiate a photon.  
Since the photon couples directly to the quark, $D_{q \to \gamma}$ is naively 
of ${\cal O}(\alpha)$ while the gluon can only couple to the photon via 
a quark and $D_{g \to \gamma}$ can be considered 
to be of ${\cal O}(\alpha\alpha_{s})$.
This simplistic argument is supported by models of the fragmentation
function \cite{bfg,grv} which suggest that 
gluon fragmentation is a much smaller effect than quark fragmentation.  
In $e^+e^-$ annihilation,
the production of gluons is suppressed by a power
of $\alpha_{s}$ compared to the production of quarks.
As a result, 
the contribution from gluon fragmentation     
is of ${\cal O}(\alpha\alpha_{s}^2)$ and should be neglected
in a consistent evaluation of this rate at fixed order up to 
${\cal O}(\alpha\alpha_{s})$. 

In this fixed order approach the inclusive photon production cross section 
can be expressed as a function of the fraction of the beam energy
carried by the photon $x_\gamma = 2 E_\gamma/\sqrt{s}$.
At ${\cal O}(\alpha)$  it takes the following form,
\begin{equation}
\frac{1}{\sigma_{0}}\frac {{\rm d} \sigma^{LO}}{{\rm d}x_{\gamma}} =
\sum_{q=1}^{2N_{F}}\;\left(
\aqed
C^{(0)}_{\gamma}(x_{\gamma},\mu_{F}) \,+\;
D_{q\to \gamma}(x_{\gamma},\mu_{F})\right), 
\label{eq:siginclo}
\end{equation}
while at ${\cal O}(\alpha \alpha_{s})$ it is given by,
\begin{equation}
\frac{1}{\sigma_{0}}\frac {{\rm d} \sigma^{NLO}}{{\rm d}x_{\gamma}}
=
\frac{1}{\sigma_0}\frac{{\rm d}\sigma^{LO}}{{\rm d}x_{\gamma}}
+ \as
\sum_{q=1}^{2N_F}\left(
\aqed
C^{(1)}_{\gamma}(x_{\gamma},\mu_{F})+ 
C_{q}^{(0)}
\otimes D_{q\to \gamma}(x_{\gamma},\mu_{F})\right).
\label{eq:sigincnlo}
\end{equation}
Here, $\sigma_{0}$ is the two-particle cross section, $N_{F}$ stands for 
the number of flavours. 
We have moreover represented the convolution in a compact way,
\begin{equation}
\int^1_x \frac{dt}{t}A(t)B\left(\frac{x}{t}\right)
= A\otimes B (x).
\end{equation}

\begin{figure}[t]
\vspace{8cm}
\begin{center}
~\includegraphics{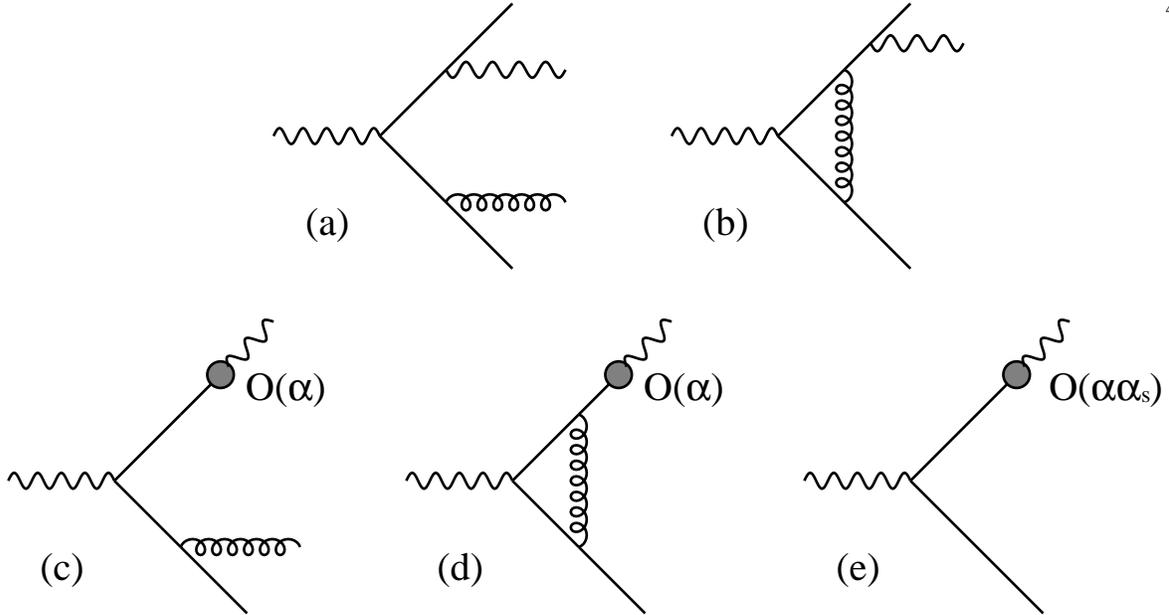}
\caption{Parton level subprocesses contributing to 
single photon production at ${\cal O}(\alpha\alpha_s)$.} 
\label{fig:class}
\end{center}
\end{figure}

The hard scattering coefficient functions $C_{i}^{(n)}$ appearing 
in these equations are defined as follows.
$C^{(0)}_{\gamma}$ is the coefficient function corresponding to 
the lowest order process $e^+e^- \to q \bar{q} \gamma$.
It is defined after the leading quark-photon singularity 
has been subtracted and factorized in the bare quark-to-photon 
fragmentation function. In the $\overline{MS}$ scheme it is given by \cite{kt},
\begin{equation}
C^{(0)}_{\gamma}(x_{\gamma},\mu_{F})=\Pqpzero(x_{\gamma})\,
\ln\left(\frac{s(1-x_{\gamma})x_{\gamma}^2}{\mu_{F}^2}\right),
\label{eq:Cgamma}
\end{equation}
where $\Pqpzero(x)$ is the $\e \to 0$ part
of the the lowest order splitting function in $(4-2\e)$-dimensions \cite{AP},
\begin{equation}
\Pqp(x) = \frac{1+(1-x)^2-\e x^2}{x}.
\label{eq:pqp}
\end{equation}
The (finite) next-to-leading order coefficient function
$C^{(1)}_{\gamma}$ can be obtained numerically 
after the next-to-leading quark-photon singularity has been subtracted.    
More precisely, $C^{(1)}_{\gamma}$ is obtained after summing 
all contributions which are independent 
of $D_{q\to \gamma}(x_{\gamma},\mu_{F})$ 
arising from the Feynman diagrams depicted 
in Fig.~\ref{fig:class} together. 
A detailed description of the evaluation of $C^{(1)}_{\gamma}$  
in the case of the `photon' +~1 jet cross section has been given  
in \cite{paper}.
The next-to-leading coefficient function appropriate for the 
inclusive photon production can moreover be 
straightforwardly obtained from the 
next-to-leading order coefficient function relevant for the 
`photon' +~1 jet cross section  by integrating the jet-specific 
variables over the complete phase space.

The coefficient function $C_{q}^{(0)}$ is the finite part 
associated with the sum of real and virtual gluon contributions 
to the process $e^+e^- \to q \bar q$. It is straightforward to evaluate, 
and can be found for example in \cite{kt},
\begin{eqnarray}
C_{q}^{(0)}(x)&=& C_{F}\left[ \left\{\frac{1+x^2}{(1-x)_{+}} 
+\frac{3}{2}\delta(1-x)\right\}\ln\left(\frac{s}{\mu_{F}^2}\right) \,+\, 
\left(\frac{2}{3}\pi^2 -\frac{9}{2}\right)\delta(1-x) \right. \nonumber\\
& & \left. + 2 \ln \left(x \frac{1+x^2}{1-x}\right)
\,+\, (1+x^2)\left(\frac{\ln(1-x)}{1-x}\right)_{+}
-\frac{3}{2}\left(\frac{1}{1-x}\right)_{+} -\frac{3}{2}x +\frac{5}{2} \right],
\end{eqnarray}
where 
$C_{F}$ the Casimir operator is 
given by, $C_F = (N^2-1)/2N$. This color factor 
is also  implicitly included in the next-to-leading order 
coefficient functions $C^{(1)}_{\gamma}$ defined above.

As motivated in \cite{letter}, within the fixed order approach
the process-independent quark-to-photon fragmentation function 
$D_{q\to \gamma}(x,\mu_{F})$ appearing in 
eq.~(\ref{eq:siginclo}) and eq.~(\ref{eq:sigincnlo}) respectively
at leading and next-to-leading order, 
satisfies an exact (up to the order under consideration) 
evolution equation. At next-to-leading order (${\cal O}(\alpha \alpha_{s})$) 
this equation reads,
\begin{equation}
\frac{\partial \Dq(x,\mu_{F})}
{\partial \ln(\mu_{F}^2)}=
\aqed \Pqpzero(x) \,+\,\aqed \as \Pqpone(x)
+\as \Pqqzero\otimes \Dq(x,\mu_{F}).
\label{eq:evolnlo}
\end{equation}
$\Pqqzero$ and $\Pqpone$ are respectively the lowest order quark-to-quark 
and the next-to-leading order quark-to-photon universal splitting 
functions \cite{AP,curci,rijken},
\begin{eqnarray}
\Pqqzero(x) &=& 
C_{F} \left[\frac{1+x^2}{(1-x)_+}+\frac{3}{2}\delta(1-x)\right],\\
\Pqpone(x)&=& C_{F} \left[ 
-\frac{1}{2} +\frac{9}{2}x + \left(-8 +\frac{1}{2}x \right) \ln x 
+2x\ln(1-x)+\left(1-\frac{1}{2}x\right)\ln^2 x \right.\nonumber\\
& &  \left.
+\left (\ln^2(1-x)+4 \ln x \ln(1-x) +8\Li_2(1-x) -\frac{4}{3}\pi^2\right)
\Pqpzero(x)\right] .
\end{eqnarray}

The quark-to-photon fragmentation function 
satisfying the evolution equation reads, 
\begin{eqnarray}
\Dq^{(NLO)}(x,\mu_{F})&=&
\Dq^{np}(x,\mu_{0}) + \aqed
\ln\left(\frac{\mu^2_{F}}{\mu_{0}^2}\right) \Pqpzero (x)
\nonumber \\
&+&\aqed \as \ln \left(\frac{\mu^2_{F}}{\mu_{0}^2}\right)
\Pqpone (x)
\nonumber\\
&+&\frac{1}{2}\aqed\as
\ln^2 \left(\frac{\mu^2_{F}}{\mu_{0}^2}\right) 
\Pqqzero\otimes \Pqpzero(x)\nonumber \\
&+&
\as
\ln \left(\frac{\mu^2_{F}}{\mu_{0}^2}\right) 
\Pqqzero\otimes\Dq^{np}(x,\mu_{0}).
\label{eq:Dnlo}
\end{eqnarray}
This solution has some interesting properties.
First, 
it is exact at the order of the calculation i.e. ${\cal O}(\alpha \alpha_s)$, 
and yields no terms of higher orders. 
The inclusive rate with this solution implemented is therefore 
independent of the choice of the factorization scale $\mu_{F}$.
The exact lowest order (${\cal O}(\alpha)$)
evolution equation and its solution $\Dq^{(LO)}(x,\mu_{F})$  
are naturally contained in eq.~(\ref{eq:evolnlo}) and eq.~(\ref{eq:Dnlo})
respectively; they can be obtained by dropping all 
terms proportional to $\alpha_{s}$ in these equations.

In eq.~(\ref{eq:Dnlo}), 
all a priori unknown non-perturbative contributions associated with the 
fragmentation function are contained in  $\Dq^{np}(x,\mu_{0})$ 
which has to be determined from the data. 
This non-perturbative input has been extracted 
from the ALEPH `photon' +~1 jet data \cite{aleph} for $\ycut=0.06$ 
and $z > 0.7$.
At lowest order, we have \cite{aleph},
\begin{equation}
\Dq^{np(LO)}(x,\mu_{0})=\aqed 
\left(-\Pqpzero(x)\ln(1-x)^2 - 13.26\right),
\label{eq:fitlo}
\end{equation}
with $\mu_{0}=0.14$~GeV while at next-to-leading order \cite{letter},
\begin{equation} 
\Dq^{np(NLO)}(x,\mu_{0})=\aqed 
\left(-\Pqpzero(x)\ln(1-x)^2 \;+\,20.8\,(1-x)
-11.07\right),
\label{eq:fitnlo1}
\end{equation}
where $\mu_{0}=0.64$~GeV and for $\alpha_s(M_Z) = 0.124$.

However, in the `photon' +~1 jet data, the process independent 
fragmentation function is extracted as a function of $z$,
the fraction of the `photon' jet momentum carried by the photon.
This is in general different from the variable relevant for the inclusive rate 
which is $x_{\gamma}$, 
the fraction of the beam momentum carried by the photon.
To see this, let us consider the lowest order process 
$e^+e^- \to q \bar{q} \gamma$, where the photon is emitted by the quark.
For this process the two variables $x_{\gamma}$ and $z$ 
are defined as follows,
\begin{equation}
x_{\gamma}=\frac{2E_{\gamma}}{\sqrt{s}}=1-\yqqb =\yqg + \yqbg, \hspace{1cm}
z=\frac{E_{\gamma}}{E_{\gamma} + E_{q}} =\frac{\yqg + \yqbg}{1 +\yqg}
\label{eq:zxdef}
\end{equation}
where  
$E_{\gamma}$ and $E_{q}$ are respectively 
the energies carried by the photon and the quark in the event 
and the dimensionless invariants $y_{ij} = (p_i+p_j)^2/s$.
Over most of phase space, the two variables are clearly different
and results derived for the `photon' +~1 jet cross section
should in principle not be transferable to the inclusive rate.
However, the non perturbative fragmentation 
effects are associated with the emission of a photon collinear 
to the quark and  
arise in the boundary region of the phase space where $\yqg \to 0$.
This is precisely where 
the definition of the two variables $z$ and $x_{\gamma}$ coincide.   
Furthermore, the collinear photon and 
quark region of the inclusive three parton phase space,
is the same as the collinear phase space restricted to the `photon' jet region,
so that the contribution to the inclusive rate 
or to `photon' + 1~jet rate from the collinear region is the same
\cite{bfg, andrew},
\begin{equation}
\frac{1}{\sigma_{0}}\frac{{\rm d}\sigma^{col}}{{\rm d}x}=
\aqed \left(\Pqpzero(x)\ln \left((1-x) x\right ) +x \right)\equiv 
C_{\gamma}^{(col)}(x)
\label{eq:sigcol}
\end{equation}  
where $x$ can stand either for $x_{\gamma}$ or for $z$.
As a consequence, the quark-to-photon fragmentation function obtained 
in the context of the calculation of the `photon' + 1~jet rate is process independent,
and can be used directly to estimate the inclusive photon rate.
This statement is in contrast with claims made in the 
literature \cite{opal,bfg}.

\begin{figure}[t]
\begin{center}
~ \epsfig{file=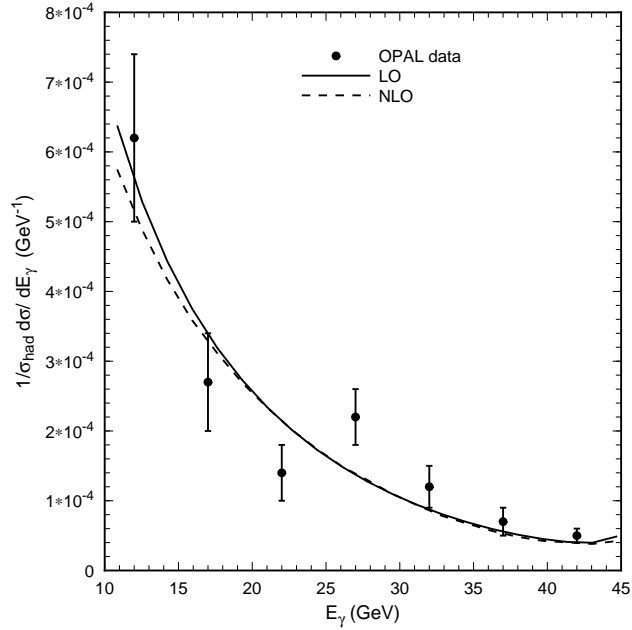,width=9cm}
\caption{The inclusive photon energy distribution 
normalized to the hadronic cross section as measured by the OPAL collaboration
compared with 
the LO (${\cal O}(\alpha)$) and NLO (${\cal O}(\alpha \alpha_s)$)
calculations including the 
quark-to-photon fragmentation function 
determined at the appropriate order from the ALEPH `photon' + 1~jet data.}
\label{fig:inclusive}
\end{center}
\end{figure} 

We therefore show our predictions for the inclusive photon energy distribution 
at both ${\cal O}(\alpha)$ and 
${\cal O}(\alpha\alpha_s)$ in Fig.~\ref{fig:inclusive}
using the fragmentation function extracted from the `photon' + 1~jet data.
For comparison we also show the measurements of the OPAL Collaboration
\cite{opal}.
We note that OPAL was able to identify photons with energies 
as little as 10~GeV, corresponding to $x_{\gamma}> 0.2$.
We see good agreement with the data for both the leading 
and next-to-leading order theoretical predictions.
It is also apparent that the next-to-leading order corrections 
to the inclusive rate are of reasonable size 
indicating that the results obtained following this fixed order 
approach are perturbatively stable.
The agreement between our predictions and the OPAL data 
is quite remarkable for another reason.
The phase space 
relevant for the OPAL data by far exceeds that used 
to determine the fragmentation function from the ALEPH `photon' +~1 jet data.
Let us quantify this statement by examining
the size of the different contributions entering in the inclusive rate 
at lowest order.

\begin{figure}[t]
\begin{center}
~ \epsfig{file=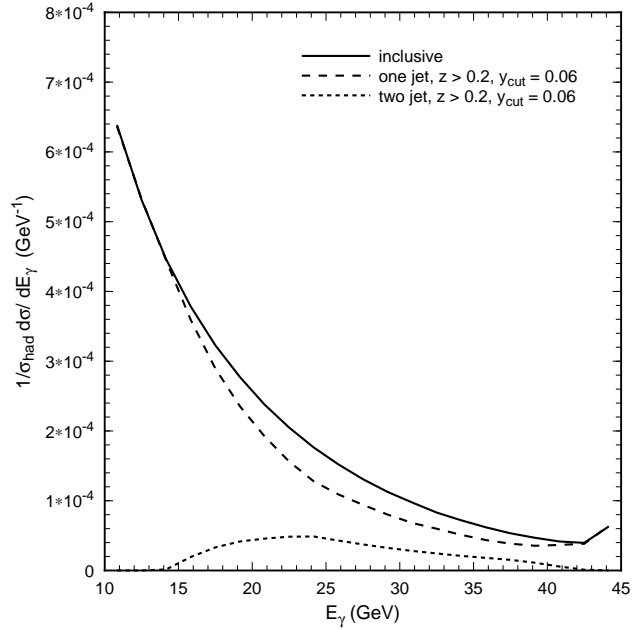,width=9cm}
\caption{The ${\cal O}(\alpha)$ inclusive photon energy spectrum (solid)
broken down into photon + $1$~jet (dashed) and
photon + $2$~jet (short-dashed) contributions 
for $z>0.2$ and for  $\ycut=0.06$.}
\label{fig:plot1}
\end{center}
\end{figure}

At lowest order, at most three jets can be formed with one of them being 
denoted as `photon' jet if it contains a photon.
To define `photon' + $n$ jet configurations,
we use the Durham algorithm \cite{durham}.
Fig.~\ref{fig:plot1} shows the relative importance for the inclusive 
rate of the `photon' +~1 jet and `photon' +~2 jet cross sections 
for $z > 0.2$ with $\ycut=0.06$.
The dominant contribution comes from `photon' + 1~jet events, precisely those used
to determine the fragmentation function.   However, the ALEPH data
used in the fit lies entirely above $E_\gamma = 32$~GeV 
(corresponding to $z > 0.7$), and the 
inclusive photon rate for smaller $E_\gamma$ should be viewed as a prediction
within the fixed order approach.
At the very least, the good agreement with the inclusive photon data
 provides some vindication for the functional form
of the fragmentation function used in fitting the high $z$ ALEPH data.

\begin{figure}[t]
\begin{center}
~ \epsfig{file=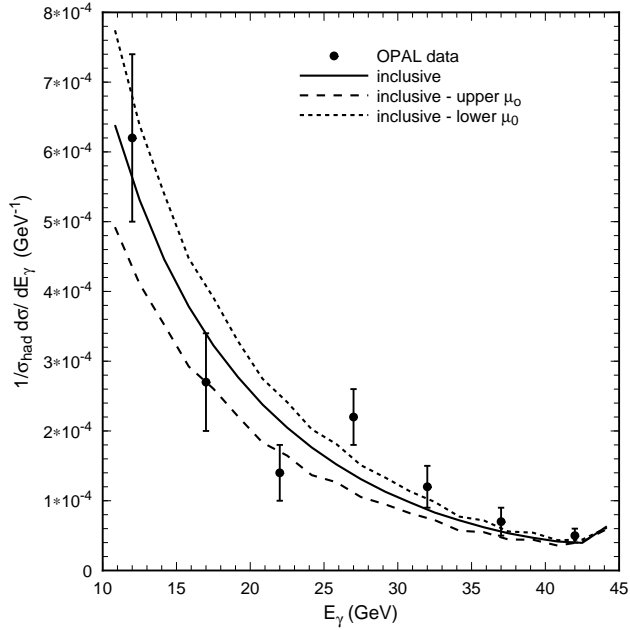,width=9cm}
\caption{The ${\cal O}(\alpha)$ inclusive photon energy
spectrum calculated using the
range of uncertainty on $\mu_0$ obtained
from the ALEPH fit to the `photon' +~1 jet data at ${\cal O}(\alpha)$. }
\label{fig:plot2}
\end{center}
\end{figure}

In order to quantify the uncertainty of our theoretical prediction 
for the inclusive rate, 
we have reevaluated it for the two extreme values allowed 
for $\mu_{0}$ as determined by ALEPH \cite{aleph}, 
\begin{equation}
\mu_{0}=0.14^{+0.30}_{-0.09}
\end{equation}
The results of this calculation together with the inclusive rate for $\mu_{0}$
being equal to its central value are depicted in Fig.~\ref{fig:plot2}.
From this figure it appears clearly that the inclusive rate 
is sensitive to the choice of the initial scale $\mu_{0}$ but also 
that the present OPAL data are too poor to enable us to reduce 
the range of uncertainty of $\mu_{0}$.

The OPAL Collaboration has compared their results with the 
estimates of~\cite{bfg, grv,owens}. They find reasonable agreement 
when the factorization scale $\mu_F$ is chosen to be equal to $M_Z$, 
although the present data were unable to discriminate between the models. 
In the following sections of this paper, we shall describe 
the different ways with which these results were obtained in some detail.
As we will see, a common feature of all these model estimates 
is the resummation of leading $(\alpha_s^n \ln^{n+1} \mu_F^2)$ 
and subleading  $(\alpha_s^n \ln^n \mu_F^2)$
logarithms of the mass factorization scale $\mu_F$ to all orders in 
the strong coupling constant.
A priori therefore, it might seem surprising that the inclusive rate 
obtained in a fixed order approach as described in this section 
and as depicted in Fig.~\ref{fig:inclusive} and the inclusive rate evaluated 
in a fundamentally different approach where the $\ln(\mu_F^2)$ are resummed
\cite{opal},
yield equally good predictions for this rate when compared to the OPAL data.
In the remainder of this paper, we shall examine more closely 
these two different approaches and one of the major aims
of this study will be to understand why these two different 
formalisms yield similar results for the inclusive rate.

\section{The conventional determination of $D_{q\to \gamma}(x,\mu_{F})$}
\setcounter{equation}{0}
\label{sec:conventional}

In the first part of this section we shall describe how 
the quark-to-photon fragmentation function is obtained in 
the so-called conventional approach. In particular we shall describe 
how the leading logarithmic (LL) and beyond-leading logarithmic (BLL)  
fragmentation functions are defined within this approach.
Analytic expansions of these functions will be given in the second  
part of this section. 

\subsection{The conventional approach}
\label{subsec:conventional}
In the conventional approach, the parton-to-photon fragmentation function
$D_{i\to \gamma}$ satisfies an all order inhomogeneous
evolution equation \cite{AP}
\begin{equation}
\frac{\partial D_{i\to \gamma}(x,\mu_F^2)}
{\partial \ln(\mu_{F}^2)}=
\left(\frac{\alpha e_{i}^2}{2\pi}\right) K_{i \gamma} + 
P_{i j} \otimes D_{j\to \gamma}  
\label{eq:ap1}
\end{equation} 
where $K_{i \gamma}(x,\mu_F^2)$ and $P_{ij}(x,\mu_F^2)$ 
are respectively the generalized photon-parton and purely partonic kernels.
Usually, these equations can be diagonalized in 
terms of the singlet and non-singlet
quark fragmentation functions as well as the gluon fragmentation function.
In the following, we shall discuss the global features of the solutions 
of these evolution equations and 
perform an expansion in powers of $\alpha_{s}$ of these solutions.
For this purpose, several simplifications can be consistently made.

As we already mentioned previously, 
it turns out that the gluon-to-photon fragmentation function is by orders 
of magnitude smaller than the quark-to-photon fragmentation functions.
As can be seen in Fig.~\ref{fig:gluon}, 
the inclusion of contributions involving the 
gluon-to-photon fragmentation function in the evaluation of the 
photon production cross sections leads only to negligible changes 
in the result, particularly at large $x$.
It is therefore legitimate when first examining the evolution equations
to ignore it and we will do so throughout this section.
A further simplification is made by ignoring the possible transitions 
between different quark flavours, which yield a small contribution 
to $\alpha^{2}_{s}P^{(1)}_{q \to q}$.
As a direct consequence of these considerations, the flavour singlet and 
non-singlet quark-to-photon fragmentation functions become equal to a unique 
fragmentation function $\Dq$ which satisfies 
an evolution equation given by,
\begin{equation}
\frac{\partial \Dq(x,\mu_F^2)}
{\partial \ln(\mu_{F}^2)}=\aqed P_{q \gamma} + P_{qq}\otimes \Dq.
\label{eq:ap2}
\end{equation}
The generalized splitting functions 
$P_{i \gamma}$ and $P_{ij}$ have a perturbative expansion 
in the strong coupling $\alpha_{s}(\mu_F^2)$. 
Explicit forms for the leading and next-to-leading 
splitting functions appearing in these expansions can be found 
in \cite{curci}. 
In particular, 
\begin{eqnarray}
P_{q \gamma}(x,\mu_F^2)&=& \Pqpzero(x) + \as \Pqpone(x) + \cdots,
\nonumber \\
P_{qq}(x,\mu_F^2)&=&\as \Pqqzero(x)   +  \cdots ,
\label{eq:Pdef}
\end{eqnarray}
with $\Pqpzero$, $\Pqpone$ and $\Pqqzero$ being 
the leading and next-to-leading order quark-to-photon and 
the leading order quark-to-quark splitting functions 
encountered in Section~\ref{sec:ourinclusive}.
The resulting evolution eq.~(\ref{eq:ap2}) 
has a similar form to  
the next-to-leading order evolution 
valid in the fixed order approach, eq.~(\ref{eq:evolnlo}).
However, unlike in 
eq.~(\ref{eq:evolnlo}), the strong coupling $\alpha_{s}$
is now a function of the factorization scale. As usual,
the running  of the strong coupling, 
$\alpha_s$, is determined by the
beta function \cite{beta},
\begin{equation}
\frac{\partial \alpha_s}
{\partial \ln(\mu_{F}^2)}=-\alpha_s^2\frac{\beta_0}{2\pi}
\left(1 + \beta_{1} \frac{\alpha_s}{2\pi}  +\cdots \right),
\label{eq:beta}
\end{equation}
where  $\beta_{0}=(33- 2N_{f})/6$ and $\beta_{1}=(306- 38 N_{f})/6\beta_0$.
  
As in our fixed order approach, 
the full solution of the inhomogeneous evolution equation is given by  
the sum of two contributions; a pointlike (or perturbative) part $\Dq^{pl}$
which is a solution of the inhomogeneous equation (\ref{eq:ap2}) 
and a hadronic (or non-perturbative) part $\Dq^{had}$ 
which is the solution of the corresponding homogeneous equation.

Approximate solutions of the evolution equations 
are commonly obtained as follows \cite{bfg,grv}.
First an analytic solution in moment space is 
obtained in the leading logarithm (LL) or beyond leading logarithm (BLL) 
approximations.
These are then inverted numerically to give 
the fragmentation function in $x$-space.
At LL only terms of the form $(\alpha_s^n \ln^{n+1} \mu_F^2)$ 
are kept while at BLL both leading
$(\alpha_s^n \ln^{n+1} \mu_F^2)$ and subleading  $(\alpha_s^n \ln^n \mu_F^2)$
logarithms of the mass factorization scale $\mu_F$ are resummed 
to all orders in the strong coupling $\alpha_{s}$.
The strong coupling itself is obtained by integrating 
eq.~(\ref{eq:beta})
and retaining only the first term in the LL case, while
keeping both terms at BLL. 
We shall examine two approximate solutions 
of the evolution equation (\ref{eq:ap2}) more closely below.

\subsubsection{The pointlike part of $\Dq$}
\label{subsec:pointlike}
Let us first concentrate on the pointlike 
part of the fragmentation function.
The moments of the fragmentation function are defined as,
\begin{equation}
\Dq(n,\mu_F)=\int_{0}^{1}{\rm d}z z^{n-1} \Dq(z,\mu_F).
\end{equation}
In moment space, the leading logarithmic solution takes the form 
\cite{bfg,grv2},
\begin{equation}
\Dq^{pl,LL}(n,\mu_F)=\aqed \frac{2 \pi}{\alpha_{s}(\mu_{F}^2)}\,a(n)
\left [1-\left( \frac{\alpha_{s}(\mu_{F}^2)}{\alpha_{s}(\mu_{0}^2)}\right)^
{1-\frac{\Pqqzero(n)}{\beta_{0}}} \right ]
\label{eq:ll}
\end{equation}
while the beyond leading logarithmic solution reads \cite{bfg,grv2},
\begin{eqnarray}
\Dq^{pl,BLL}(n,\mu_F)&=& \aqed
\left \{\left [ \frac{2 \pi}{\alpha_{s}(\mu_{F}^2)}\,a(n) \,+\,b'(n) \right] \;
\left [1-\left( \frac{\alpha_{s}(\mu_{F}^2)}{\alpha_{s}(\mu_{0}^2)}\right)^
{1- \frac{\Pqqzero(n)}{\beta_{0}}} \right ] \right.
\nonumber \\
&& \left.  + b(n)
\left [1-\left( \frac{\alpha_{s}(\mu_{F}^2)}{\alpha_{s}(\mu_{0}^2)}\right)^
{- \frac{\Pqqzero(n)}{\beta_{0}}} \right ] \right \},
\label{eq:bll}
\end{eqnarray}
with $\Pqqzero(n)$
denoting 
the moments of the leading order 
quark-to-quark splitting function.

Independently of the precise definitions of the functions $a$, $b$ and $b'$ 
which we will come to next, both LL and BLL solutions have an asymptotic 
behaviour given by,
\begin{equation}
\Dq^{asympt}(n,\mu_F)=\aqed \frac{2 \pi}{\alpha_{s}(\mu_{F}^2)}\,a(n).
\label{eq:asympt}
\end{equation}
This asymptotic form lends support to
the common assumption that the
quark-to-photon fragmentation function $\Dq$ is   
${\cal O}\left( \alpha/\alpha_{s} \right)$. 
This assumption is in contrast with 
that adopted in the fixed order approach (cf. Section~\ref{sec:ourinclusive}) 
where the quark-to-photon fragmentation function is  ${\cal O}(\alpha)$.
It can lead 
to significant differences in the respective expressions 
of the one-photon production cross sections. 
We shall study these discrepancies 
more closely in Section~\ref{sec:comparison}.

The functions $a$, $b'$ and $b$ are functions 
of the leading and next-to-leading 
quark-to-quark and quark-to-photon splitting functions in moment space.
They are given by \cite{bfg,grv2},
\begin{eqnarray}
a(n)&=&\frac{1}{1-\frac{\Pqqzero(n)}{\beta_{0}}}\,
\frac{\Pqpzero(n)}{\beta_{0}}
\nonumber \\
b'(n)&=& -\frac{2}{\beta_{0}} 
\left [\Pqqone(n) 
-\beta_{1} \Pqqzero(n)\right] a(n)
\nonumber \\
b(n)&=&-\frac{1}{\Pqqone(n)}\left [
\Pqpone(n)
- \beta_{1}\Pqpzero(n)
+\frac{\Pqpzero(n)}{\beta_{0}}
\left (\Pqqone(n) -\beta_{1}
\Pqqzero(n) \right) \right ].
\label{eq:ab'bdef}
\end{eqnarray}
Here $P_{a\to b}^{(m)}(n)$
denote the moments of the $m$th order $a \to b$ splitting functions.
They can be found in \cite{grv}.

We will explicitly use these definitions in Sec.~\ref{subsec:expansion}
to estimate the difference between the numerical LL 
and BLL solutions in $x$-space 
obtained by inverting eqs.~(\ref{eq:ll}) and (\ref{eq:bll})
\cite{bfg,grv} and analytic expressions obtained making a Taylor 
expansion in $\alpha_s$ of these LL and BLL solutions. 

\subsubsection{The hadronic part of $\Dq$}
\label{subsec:hadronic}
The hadronic part of the quark-to-photon fragmentation 
function is a solution of the homogeneous evolution equation
(eq.~(\ref{eq:ap2}) with $P_{q \gamma}=0$).
As for the solution of the inhomogeneous evolution equation,  
we can obtain solutions in the LL 
and BLL approximations defined above.
In moment space we have,
\begin{equation}
\Dq^{had,LL}(n,\mu_F)=\left 
[\left( \frac{\alpha_{s}(\mu_{F}^2)}{\alpha_{s}(\mu_{0}^2)}\right)^
{-\frac{\Pqqzero(n)}{\beta_{0}}} \right ]\;\Dq^{had,LL}(n,\mu_0)
\;+\; {\cal O}(\alpha_{s})
\label{eq:dhadll}
\end{equation}
and \cite{grv2},
\begin{eqnarray}
\Dq^{had,BLL}(n,\mu_F)&=&
\left 
[\left( \frac{\alpha_{s}(\mu_{F}^2)}{\alpha_{s}(\mu_{0}^2)}\right)^
{-\frac{\Pqqzero(n)}{\beta_{0}}} \right ]\;\Dq^{had,BLL}(n,\mu_0)
\nonumber \\
& & \times
\left \{ 1-\frac{\alpha_{s}(\mu_{F}^2) -\alpha_{s}(\mu_{0}^2)}
{2\pi \beta_{0}}\,\left ( \Pqqone(n) -\beta_1
\Pqqzero(n) \right) \right \}\,+\,{\cal O}(\alpha_{s}^2).
\nonumber\\
\label{eq:dhadbll}
\end{eqnarray}

To obtain the pointlike LL and BLL solutions of 
evolution equations (eqs.~(\ref{eq:ll}) and (\ref{eq:bll})), 
in any case, one needs to specify the non-perturbative 
input $\Dq^{np}(n,\mu_0)$, 
which in both the conventional \cite{bfg,grv} and fixed 
order approaches \cite{letter} is proportional to $\alpha$.
Once this initial fragmentation function is chosen, within the conventional
approach there are two different ways adopted in the literature 
to define the complete fragmentation function.

One way is to consider the complete solution in a given approximation 
to be obtained as the sum of the pointlike and hadronic solutions 
within that approximation.
In particular, the  full BLL solution is obtained by adding the BLL 
pointlike and hadronic parts together.
In this approach, which is 
adopted by Gl\"{u}ck, Reya and Vogt (GRV) in \cite{grv}
 the full solution can be obtained by iteration. However, 
when solving the evolution equation in moment space, 
one inevitably includes terms which are 
beyond the order considered 
(as can be seen in eqs.~(\ref{eq:dhadll}) and  (\ref{eq:dhadbll})). 
When the inversion into $x$-space is performed, 
one obtains spurious terms which can lead to significant contributions
and have to be systematically omitted \cite{grv}.

An alternative way to construct the full BLL 
fragmentation function, adopted by Bourhis, Fontannaz and Guillet (BFG) 
in \cite{bfg}, is to associate the BLL pointlike part with the LL 
hadronic part. 
From eq.~(\ref{eq:Dnlo}), it appears that the treatment 
of the hadronic part of the fragmentation function in this second conventional 
approach is conceptually closer to its treatment in the fixed order 
approach. 

At this stage it is important to mention that,
in either of the two conventional approaches used to determine 
the quark-to-photon fragmentation function $\Dq(x,\mu_{F})$ described above, 
the hadronic input associated with the LL pointlike part turns out to be 
negligible and can be described 
by a VMD model \cite{bfg,grv}. However, the hadronic input associated 
with the BLL pointlike part {\em is} sizeable  
and cannot be purely described 
by such a model anymore. We will come back to this important point and to 
the possible forms of this hadronic 
input in some detail in Section~\ref{sec:comparison}.

\subsection{Analytic expansion of $D^{pl}_{q\to \gamma}(x,\mu_{F})$}
\label{subsec:expansion}
In this subsection we shall concentrate on the pointlike part 
of the fragmentation function and more precisely on obtaining 
an analytic expression for it by making a series expansion in $\alpha_s$.
As already mentioned, the LL and BLL resummed expressions in $x$-space 
of the pointlike fragmentation function can be obtained by inverting 
numerically eqs.~(\ref{eq:ll}) and (\ref{eq:bll}). 
Approximations of these resummed solutions in $x$-space can 
however be obtained analytically. 
First, one expands the expressions for the resummed fragmentation functions 
in moment space as a series in $\alpha_{s}$, up to a given order. 
The truncated 
series can then easily be inverted analytically 
to yield an approximate expression for $\Dq^{pl}$ in $x$-space. 
These expanded expressions of the pointlike quark-to-photon 
fragmentation function can then be compared  
with the LL and BLL resummed expressions of $\Dq^{pl}$ 
which are only known numerically.   

More precisely,
an analytic expansion (up to ${\cal O}(\alpha \alpha_{s})$) 
of the LL expression for the fragmentation function is obtained as follows.
First, the LL expression for $\alpha_{s}(\mu_{0}^2)$
truncated at order $\alpha^{3}_{s}$ reads,
\begin{eqnarray}
\alpha_{s}^{(LL)}(\mu_{0}^2)&=&\alpha_{s}(\mu_{F}^2)
\left[ 1-\left(\frac{\alpha_{s}(\mu_{F}^2)\beta_{0}}{2 \pi}\right) 
\ln\left(\frac{\mu_{F}^2}{\mu_{0}^2}\right)
\right]^{-1}
\nonumber \\
&=& \alpha_{s}(\mu_{F}^2) \left[
1+\left(\frac{\alpha_{s}(\mu_{F}^2)\beta_{0}}{2 \pi}\right) 
\ln\left(\frac{\mu_{F}^2}{\mu_{0}^2}\right)
+\left(\frac{\alpha_{s}(\mu_{F}^2)\beta_{0}}{2 \pi}\right)^2 
\ln^2\left(\frac{\mu_{F}^2}{\mu_{0}^2}\right)\right] 
+ O(\alpha^3_{s}).
\nonumber\\
\label{eq:alphasll}
\end{eqnarray}
Inserting this expression into eq.~(\ref{eq:ll}) 
and expanding in series up to order $\alpha_{s}(\mu_{F}^2)$ one obtains, 
\begin{eqnarray}
\Dq^{pl,LL (exp.)}(n,\mu_F)&=&
\aqed \ln\left(\frac{\mu^2_{F}}{\mu_{0}^2}\right) \Pqpzero(n) 
+ \frac{1}{2}\aqed\as\ln^2\left(\frac{\mu^2_{F}}{\mu_{0}^2}\right)
\Pqqzero(n)\Pqpzero(n)
 \nonumber\\
& & +O(\alpha^2_{s}),
\end{eqnarray} 
which can be trivially inverted to yield the 
expansion of the LL pointlike 
fragmentation function in $x$-space,  
\begin{eqnarray}
\Dq^{pl, LL (exp.)}(x, \mu_F)&\equiv&
\aqed \ln\left(\frac{\mu^2_{F}}{\mu_{0}^2}\right)
\Pqpzero(x)
+\frac{1}{2}
\aqed\as\ln^2\left(\frac{\mu^2_{F}}{\mu_{0}^2}\right)
\Pqqzero \otimes \Pqpzero(x)
 \nonumber\\
& & +O(\alpha^2_{s}).
\label{eq:llexp}
\end{eqnarray}

Similarly, an expansion (up to ${\cal O}(\alpha \alpha_{s})$)
of the BLL pointlike fragmentation function is obtained by considering 
the BLL expression of $\alpha_{s}(\mu_{0}^2)$,  i.e. the expression obtained 
retaining the term proportional to $\beta_1$, which reads,
\begin{eqnarray}
\alpha_{s}^{(BLL)}(\mu_{0}^2)&=& \alpha_{s}(\mu_{F}^2)
\left[ 1-\left(\frac{\alpha_{s}(\mu_{F}^2)\beta_{0}}{2 \pi}\right)\left\{ 
\ln\left(\frac{\mu_{F}^2}{\mu_{0}^2}\right) 
+\left(\frac{\alpha_{s}(\mu_{F}^2)\beta_{1}}{2 \pi}\right)\,
\ln\left(\frac{\mu_{F}^2}{\mu_{0}^2}\right)\right\}  \right]^{-1}
\nonumber \\
&=& \alpha_{s}(\mu_{F}^2) \left[
1+\left(\frac{\alpha_{s}(\mu_{F}^2)\beta_{0}}{2 \pi}\right) \left\{
\ln\left(\frac{\mu_{F}^2}{\mu_{0}^2}\right)
+\left(\frac{\alpha_{s}(\mu_{F}^2)\beta_{1}}{2 \pi}\right) 
\ln\left(\frac{\mu_{F}^2}{\mu_{0}^2}\right) \right. \right.
\nonumber\\
&& \left. \left. \qquad
\qquad \;+\left(\frac{\alpha_{s}(\mu_{F}^2)\beta_{0}}{2 \pi}\right) 
\ln^2\left(\frac{\mu_{F}^2}{\mu_{0}^2}\right) \right \} \right]
\,+ O(\alpha^3_{s}).
\label{eq:alphasbll}
\end{eqnarray}
Inserting it into the 
resummed expression of the BLL fragmentation function 
given in eq.~(\ref{eq:bll}),
the expanded expression of the BLL pointlike fragmentation function 
in $x$-space thus reads,
\begin{equation}
\Dq^{pl,BLL (exp.)}(x,\mu_F) \equiv 
\Dq^{pl, LL (exp.)}(x, \mu_F)+\aqed 
\as\ln\left(\frac{\mu^2_{F}}{\mu_{0}^2}\right)
\Pqpone(x)
+O(\alpha^2_{s}).
\label{eq:bllexp}
\end{equation}

\begin{figure}[t]
\begin{center}
~\epsfig{file=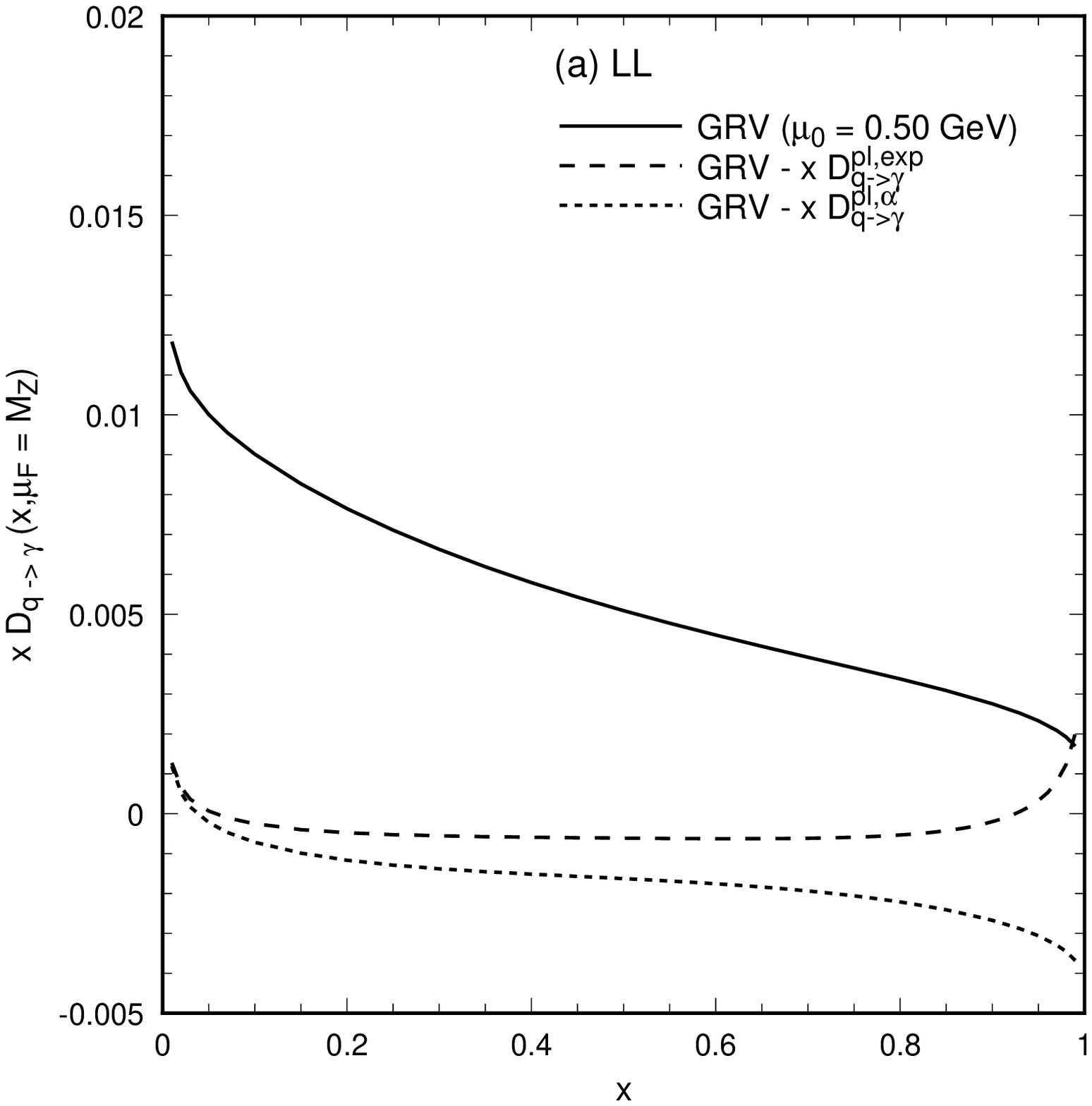, width=7cm} 
~\epsfig{file=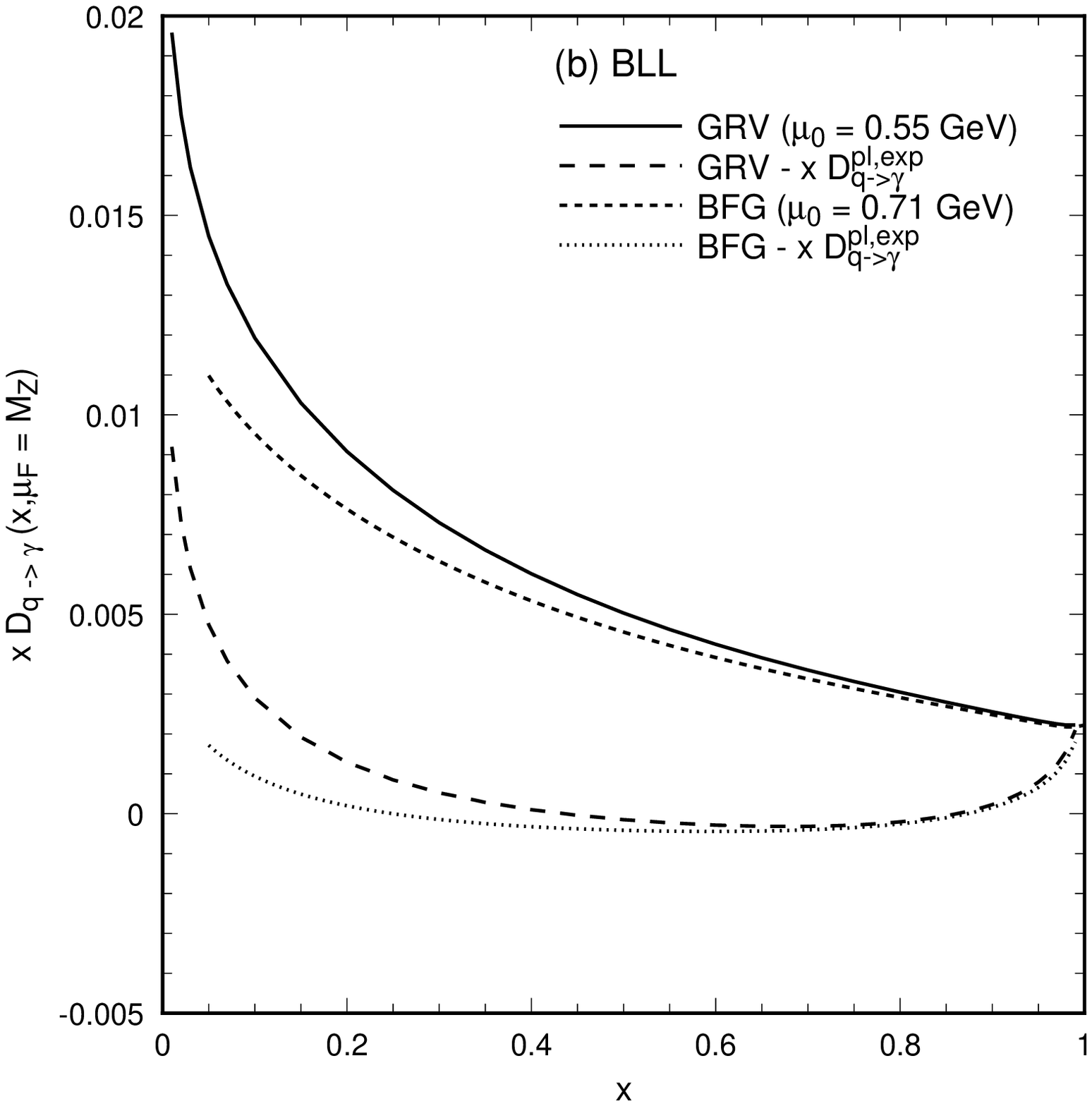, width=7cm}
\caption{
The pointlike quark-to-photon fragmentation function for 
the up-quark evaluated at $\mu_F = M_Z$
at (a) LL and (b) BLL.
In each case, we show the numerical resummed prediction 
as well as the difference with the corresponding
fixed order expansions given in eqs.~(\ref{eq:llexp}) and (\ref{eq:bllexp}).
The hadronic scale is (a) $\mu_0 = 0.50$~GeV for
LL GRV and (b)  
$\mu_0 = 0.55$~GeV for BLL GRV and $\mu_0 = 0.71$~GeV for BLL BFG.
In (a) the difference between the LL GRV result and
the leading term in the perturbative expansion eq.~(\ref{eq:alpha})
is shown short-dashed.}
\label{fig:plll}
\end{center}
\end{figure}  

The expanded expressions for the pointlike part of the 
quark-to-photon fragmentation function can 
now be compared directly 
with the numerical solutions of the LL  
and BLL resummed expressions 
for the pointlike fragmentation function
for a fixed value of $\mu_{F}$ and over the whole $x$ range.
These comparisons are shown in 
Fig.~\ref{fig:plll} 
using the parameterization of the LL and BLL fragmentation functions
given by Gl\"{u}ck, Reya and Vogt (GRV) in  \cite{grv}  and 
the BLL parameterization given 
by Bourhis, Fontannaz and Guillet (BFG) in \cite{bfg}.
In each case, we show the fragmentation function for the up-quark multiplied 
by the momentum fraction $x$. The other flavours have a similar behaviour. 
To compare the series expanded fragmentation function 
with the resummed expression, 
we also show the differences between the resummed and the expanded 
solutions given in eqs.~(\ref{eq:llexp}) and (\ref{eq:bllexp}) for the 
appropriate choices of $\mu_F$ and $\mu_0$.  That is $\mu_0 = 0.50$~GeV for
LL GRV, $\mu_0 = 0.55$~GeV for BLL GRV and $\mu_0 = 0.71$~GeV for BLL BFG.
In all cases we choose $\mu_F = M_Z$.
As a further comparison, in Fig.~\ref{fig:plll}(a)
we also show the difference between the resummed LL
fragmentation function and the ${\cal O}(\alpha)$ term of the series expansion,
\begin{equation}
\Dq^{pl,\alpha}(x,\mu_F^2) = \aqed\Pqpzero \ln\left(\frac{\mu_F^2}{\mu_0^2}
\right).
\label{eq:alpha}
\end{equation} 

Inspection of Fig.~\ref{fig:plll}(a) suggests that $\Dq^{pl,\alpha}$ 
which is the only term present in the lowest order solution
obtained in the fixed order approach is insufficient to correctly reproduce
the behaviour of the LL resummed expression.
On the other hand, we see that in the region $0.2<x<0.9$ 
the expanded expression 
of the fragmentation function  is remarkably close to the resummed 
expression for both LL and BLL pointlike solutions 
and in the BLL case for the both parameterizations (GRV or BFG).
For small $x$, $ x < 0.2$, 
there are possible large logarithms of
$x$ in addition to contributions from the gluon fragmentation function 
that are not treated correctly in the expanded result.  
At large $x$, $x> 0.9$, there is also a significant difference
between the two approaches.  
This discrepancy could be a sign that large resummation effects are present.
Or it could indicate that the presently available  
parameterizations for the resummed 
fragmentation functions are not accurate at large $x$ and particularly for
$x > 0.95$.
In fact,  this discrepancy can be traced back to the presence
of logarithms of $(1-x)$ that are explicit in the expanded result. 
These logarithms should also be present in the numerical resummed results.
However, the parameterizations are necessarily
obtained by inverting only a finite number of moments
and it is a well known problem to describe a logarithmic behaviour with a 
polynomial expansion.

As the resummed fragmentation functions 
were obtained after 
the $\ln(\mu_{F}^2)$ had been resummed, the general
agreement with the unsummed
and expanded fragmentation functions leads us to question 
the necessity of such a resummation at LEP energies.  
Moreover, this agreement in the BLL case has another important consequence.
As can be seen from eq.~(\ref{eq:bllexp}) 
the expanded expression for the BLL pointlike 
quark-to-photon-fragmentation function is also equal 
to the expression of the next-to-leading 
perturbative fragmentation function obtained in the fixed order approach 
as given in eq.~(\ref{eq:Dnlo}), where the hadronic input is neglected,
$\Dq^{np}(x,\mu_{0})=0$.  
It is therefore instructive
to implement the expanded expression for the quark-to-photon 
fragmentation function in the evaluation of observable one photon
cross sections at LEP energies.
Indeed, doing so 
will enable us to compare the results for these cross sections 
obtained in different approaches and to isolate easier 
the differences between them, a task to which we will now turn in 
Section~\ref{sec:comparison}.

\section{The cross section in the different approaches compared}
\setcounter{equation}{0}
\label{sec:comparison}

We are finally interested in comparing the inclusive 
and `photon' +~1 jet cross sections evaluated 
in the two essentially different approaches, 
at fixed order and following the conventional approach.
In Section~\ref{sec:ourinclusive}, we have described how the fragmentation function 
and the one-photon cross section are defined in the fixed order approach,
while in Section~\ref{sec:conventional} the derivation of the LL and BLL expressions of the 
fragmentation function has been discussed. These resummed 
expressions were determined within the conventional approach, 
as approximations of the solution of an all-order evolution 
equation~(\ref{eq:ap2}).
By making a Taylor expansion in the strong coupling,
analytic expressions for the LL and BLL 
pointlike solutions were also considered in Section~\ref{subsec:expansion}.
Nothing however has been said so far concerning the expressions 
for the single photon production cross section within this conventional 
formalism. We shall fulfill this task in this section.

In the following, we shall consider four different classes of expressions 
for the one-photon production cross section.
These classes will be defined depending on whether   
the resummed (LL or BLL) expressions of the quark-to-photon 
fragmentation function 
or the expanded expression of the fragmentation 
function as given by eq.~(\ref{eq:bllexp}) are used 
in the cross section. Secondly these classes will be determined 
depending on whether the direct contributions to the cross section 
are evaluated as a perturbative series in $\alpha_{s}$
up to ${\cal O}(\alpha_{s})$, or whether these direct contributions  
are evaluated by using a conventional power counting, associating 
the powers of $\alpha_{s}$ and the powers of $\ln\mu^2_{F}$ together. 
The results obtained for the 
inclusive and `photon' + 1 jet cross sections following any of these four approaches 
to evaluate the one-photon production cross section will be compared 
to the OPAL and ALEPH data in the forthcoming section.

As we will see, in the category of approaches 
using the expanded expression of the quark-to-photon fragmentation function 
in the cross section, the specification of the input function 
$D_{q \to \gamma}^{np}(x,\mu_{0})$ in a 
given factorization scheme is an important and subtle point 
which is treated  differently by GRV in \cite{grv} 
and by Bourhis, Fontannaz and Guillet BFG in  \cite{bfg}.
We shall describe the determination of this input fragmentation function 
according to either of these two groups in some detail in the second part 
of this section.

\subsection{Approaches using the resummed $D_{q\to \gamma}(x,\mu_{F})$}
\subsubsection{Direct contributions evaluated at fixed order in $\alpha_{s}$}
\label{subsubsec:resfix}

Let us first concentrate on the expression of the one-photon production cross 
section in the $\overline{{\rm MS}}$ factorization scheme obtained 
using the LL or BLL resummed expressions of the  
quark-to-photon fragmentation function while keeping 
the direct hard scattering terms building the cross section at fixed order 
in the strong coupling constant $\alpha_{s}$ up to ${\cal O}(\alpha\alpha_{s})$.
In this case the cross section takes the same form as in the 
fixed order approach described in Section~\ref{sec:ourinclusive}, 
and is described by eq.~(\ref{eq:siginclo}) 
(with $x_{\gamma}$ replaced by $x$)
at ${\cal O}(\alpha)$ and by eq.~(\ref{eq:sigincnlo}) 
at ${\cal O}(\alpha\alpha_{s})$.
Rather than the fitted forms (eqs.~(\ref{eq:fitlo}) and (\ref{eq:fitnlo1})),
at ${\cal O}(\alpha)$ the LL fragmentation function should be considered 
in eq.~(\ref{eq:siginclo}),
while at ${\cal O}(\alpha\alpha_{s})$ the BLL fragmentation function 
needs to be taken into account in eq.~(\ref{eq:sigincnlo}).
Provided, the solution of the all order evolution equation 
can be accurately determined, the cross section evaluated 
following this approach is theoretically preferred as it is 
the most complete: It includes all direct 
terms up to order $\alpha \alpha_{s}$ and all fragmentation contributions  
proportional to $(\alpha_{s}^{n}\ln^{n}\mu_{F}^{2})$ and 
$(\alpha_{s}^{n}\ln^{n+1}\mu_{F}^{2})$ at all orders.

In order to evaluate either of the single photon production rates according 
to this prescription, we simply need to replace the fragmentation functions 
defined at fixed order appearing in these leading and next-to-leading 
cross sections 
by the resummed  LL or BLL fragmentation functions 
and leave the remaining terms in the cross sections unchanged.
Consequently, by following this approach one is in principle able to test 
the universality of the LL and BLL fragmentation functions, 
in particular when these functions are employed
to evaluate the `photon' + 1 jet rate, 
an observable which was not used to determine these functions.

\subsubsection{Direct contributions conventionally evaluated}
\label{subsubsec:rescon}

Second, let us consider the expression of the cross section in the 
$\overline{{\rm MS}}$ scheme arising when one uses the
resummed LL or BLL fragmentation functions 
but when one considers the direct terms 
of the  single photon production cross section evaluated 
with the conventional power counting. These terms are obtained by  
keeping only the leading or beyond leading logarithmic terms  
of the factorization scale $\mu_{F}$ up to a given order in $\alpha_{s}$.
As it is commonly done \cite{bfg,grv}, we shall follow 
this prescription, which defines the conventional approach to obtain the LL
and BLL cross sections, and only retain terms proportional to 
 $(\alpha_{s}^{n}\ln^{n+1}\mu_{F}^2)$ at LL 
and terms of the form $(\alpha_{s}^{n}\ln^{n+1}\mu_{F}^2)$ 
and $(\alpha_{s}^{n}\ln^{n}\mu_{F}^2)$ in the BLL 
expression of the cross section.

Remembering furthermore that in this conventional approach, 
an inverse power of $\alpha_{s}$ is associated with the quark-to-photon 
fragmentation function as in eq.~(\ref{eq:asympt}), the LL and BLL expressions  
of the cross section are simply,
\begin{eqnarray}
\frac{1}{\sigma_{0}}\frac {{\rm d} \sigma^{LL}}{{\rm d}x} &=&
\sum_{q=1}^{2N_{f}}\;D_{q\to \gamma}(x,\mu_{F}), 
\nonumber\\
\frac{1}{\sigma_{0}}\frac {{\rm d} \sigma^{BLL}}{{\rm d}x}
&=&\sum_{q=1}^{2N_{f}}\;
\left(
D_{q\to \gamma}(x,\mu_{F})\,
+\as C_{q}^{(0)}
\otimes D_{q\to \gamma}(x,\mu_{F}) 
+ \aqed  C^{(0)}_{\gamma}(x,\mu_{F})\right).
\label{eq:sigbll}
\end{eqnarray}
As already mentioned before, due to the different 
power of $\alpha_{s}$ associated with the fragmentation function 
in this approach and in the fixed order approach, the terms present 
in these two equations and in the corresponding fixed order equations 
((\ref{eq:siginclo}) and (\ref{eq:sigincnlo})) differ substantially.  
As explained at length in~\cite{letter}, this conventional 
procedure of associating an inverse power of $\alpha_{s}$ 
with the fragmentation function 
is clearly appropriate when the logarithms of the factorization scale 
$\mu_{F}$ are the {\em only} potentially large logarithms but 
is theoretically less consistent when different classes of large 
logarithms can occur as in the `photon' +~1 jet cross section.

\subsection{Approaches using the expanded $D_{q\to \gamma}(x,\mu_{F})$}

In this subsection, we shall consider the formulations of the cross section 
obtained using the expanded (up to ${\cal O}(\alpha \alpha_{s})$) 
expressions of the pointlike and 
hadronic part of the fragmentation function $\Dq(x,\mu_{F})$.
In particular, we will consider  
the pointlike part of the fragmentation function  
given by the expanded expression of the resummed form 
for the BLL pointlike fragmentation function discussed
in Section~\ref{subsec:expansion} and defined in eq.~(\ref{eq:bllexp}).
An expanded expression of the hadronic part will be given below.
Finally, we shall implement the expanded expression obtained for the 
sum of pointlike and hadronic parts of the fragmentation function  
in the NLO  and BLL 
formulations of the cross section 
given by eqs.(\ref{eq:sigincnlo}) and (\ref{eq:sigbll}) respectively.

\subsubsection{Possible definitions of $D_{q\to \gamma}^{np}(x,\mu_{0})$} 

In order to obtain a definite prediction for the cross section using  
either the GRV or BFG models of the quark-to-photon 
fragmentation function, we need to know how the 
hadronic part of the fragmentation function behaves 
and in particular we need to know 
how the input function $D_{q \to \gamma}^{np}(x,\mu_{0})$ is defined.
Since within this context we have an expanded 
(up to ${\cal O}(\alpha \alpha_{s})$) form for the pointlike 
part of the fragmentation function, we here choose to consider also
an expanded expression (up to ${\cal O}(\alpha \alpha_{s})$) 
for the hadronic part of this function, given by,
\begin{equation}
\Dq^{had,(exp.)}(x,\mu_{F})=D_{q \to \gamma}^{np}(x,\mu_{0}) +
\as\ln \left(\frac{\mu^2_{F}}{\mu_{0}^2}\right)\Pqqzero\otimes 
D_{q \to \gamma}^{np}(x,\mu_{0})+ {\cal O}(\alpha \alpha_s^2).
\label{eq:Dhadexpdef}
\end{equation} 

This expression is obtained by expanding eq.~({\ref{eq:dhadbll}) 
and takes exactly the same form as the non-perturbative part 
of the fragmentation function defined in the fixed order 
approach described in Section~\ref{sec:ourinclusive}. 
Recall that in the fixed order approach, 
the input fragmentation function $D_{q \to \gamma}^{np}(x,\mu_{0})$ 
present as a boundary condition in eq.~(\ref{eq:Dhadexpdef})
was determined at each order 
by comparing the fixed order `photon' +1~jet cross section with the ALEPH data,
with leading order and next-to-leading order expressions given in equations    
(\ref{eq:fitlo}) and (\ref{eq:fitnlo1}) respectively.
Note also that the fitted   
input fragmentation function determined within this fixed order 
context is clearly non-negligible at any order in $\alpha_{s}$.

In the approaches of GRV or BFG , 
the treatment of the input fragmentation function 
$D_{q \to \gamma}^{np}(x,\mu_{0})$ is quite different.
At LL both GRV and BFG agree  
that $D_{q \to \gamma}^{np}(x,\mu_{0})$
is negligible and can be described by a vector meson dominance model (VMD) 
as explained in \cite{bfg} and \cite{grv} respectively.
However at BLL and in the $\overline{{\rm MS}}$ scheme, 
the input fragmentation function cannot be negligible 
due to the presence of the direct term $C_{\gamma}^{(0)}$ 
(see eqs.~(\ref{eq:siginclo}), (\ref{eq:sigincnlo}) or (\ref{eq:sigbll})) 
and 
cannot be described by a VMD  input alone. 
Indeed, $C_{\gamma}^{(0)}$ diverges as $x \to 1$ and would drive 
the cross section to unacceptable negative values if a VMD input 
alone is considered for the input fragmentation function.
Note that the requirement that the
cross section is positive led the authors in \cite{andrew, letter} to consider 
a term proportional to $\Pqpzero \ln(1-x)^2$ in the expression of 
$D_{q \to \gamma}^{np}(x,\mu_{0})$.
To summarize the discussion, in any resummed or fixed order approach, as soon 
as the direct term $C_{\gamma}^{(0)}$ enters the cross section,
as it does in the $\overline{{\rm MS}}$ factorization scheme, 
the input fragmentation function 
$D_{q \to \gamma}^{np}(x,\mu_{0})$ must compensate the large $x$
behaviour of $C_{\gamma}^{(0)}$.  

So far all the formulae given for the cross section 
were given in the $\overline{{\rm MS}}$ factorization scheme.
It is well known however, that the direct and fragmentation contributions 
are not unequivocally defined. 
For instance 
in a different factorization scheme, $S$
a part of $C_{\gamma}^{(0)}$ (direct term) $\Delta C_{\gamma}^{(0)} \mid _{S}$
can be absorbed 
in the fragmentation function  yielding 
new functions $\Dq(x,\mu_{F})\mid _{S}$ and 
$C_{\gamma}^{(0)}\mid_{S}(x,\mu_{F})$, so that,
\begin{equation}
\Dq(x,\mu_{F})\mid _{S}=\Dq(x,\mu_{F})
\mid _{\overline{{\rm MS}}}      
+\aqed \Delta C_{\gamma}^{(0)} \mid _{S}(x) 
\label{eq:DSdef}
\end{equation}
and,
\begin{equation}
C_{\gamma}^{(0)}\mid_{S}(x,\mu_{F})=
C_{\gamma}^{(0)}\mid _{\overline{{\rm MS}}}(x,\mu_{F}) 
- \Delta C_{\gamma}^{(0)} \mid _{S}(x).
\label{eq:CSdef}
\end{equation}
In fact only the combination of both direct and fragmentation contributions present in  
the physical cross section is factorization scheme invariant. 
The approaches adopted by GRV or BFG use this factorization
scheme ambiguity to determine 
$D_{q \to \gamma}^{np}(x,\mu_{0})$, albeit
in a slightly different way.

The GRV group choose to work within a 
factorization scheme (called $DIS_{\gamma}$) 
in which
the fragmentation input $D_{q \to \gamma}^{np}(x,\mu_{0})$
at both LL and BLL is simply given by a VMD contribution 
and is therefore negligible.
Essentially, within this new scheme the troublesome part of the direct 
contribution $C_{\gamma}^{(0)}$ appearing in eq.~(\ref{eq:sigbll}) 
is removed by absorbing it into the 
definition of the quark-to-photon fragmentation function $\Dq(x,\mu_{F})$.
More precisely, in this scheme we have,
\begin{equation}
\Delta C_{\gamma}^{(0)} \mid_{DIS_{\gamma}}(x)= \Pqpzero(x)
 \ln\left((1-x)x^2\right) -2 \left( \frac{1-x}{x}\right) ,
\end{equation}
and,
\begin{equation}
\Dq^{np}(x,\mu_{0})\mid _{DIS_{\gamma}}=\Dq^{np}(x,\mu_{0})
\mid _{\overline{{\rm MS}}}      
+\; \aqed \Delta C_{\gamma}^{(0)} \mid_{DIS_{\gamma}}(x) \sim \Dq^{VMD}.
\label{eq:DinputDIS}
\end{equation}
This transformation holds for any quark flavour.
Within this scheme, the direct term 
$C_{\gamma}^{(0)} \mid_{DIS_{\gamma}}$ is 
regular as $x\to 1$. 
The evolution equation of the quark-to-photon fragmentation function 
in the $S$-scheme is modified as well, 
\begin{equation}
\Pqpone \mid_{S}=\Pqpone \mid _{\overline{{\rm MS}}}\;
-\,\aqed \Delta C_{\gamma}^{(0)} \mid_{S}\otimes\Pqqzero.
\end{equation}
Invoking a perturbative stability
argument, this group considers the evolution equation in the 
$DIS_{\gamma}$-scheme, solves this equation in that scheme 
and then since the $\overline{{\rm MS}}$-scheme is traditionally preferred 
in the evaluation of cross sections at higher orders, 
transforms back their results to the $\overline{{\rm MS}}$-scheme  
according to eq.~(\ref{eq:DSdef}).
For illustration, the GRV up-quark fragmentation function at $\mu_F= M_Z$ 
and in both the $DIS_{\gamma}$ and $\overline{{\rm MS}}$ factorization schemes
is shown in Fig.~\ref{fig:grvDu}.

Note that, to obtain the results in the $DIS_{\gamma}$ scheme, 
we have added a term proportional to 
$\Delta C_{\gamma}^{(0)}\mid_{DIS_{\gamma}}(x)$ to the
$\overline{{\rm MS}}$ fragmentation function provided numerically by GRV.
We see that the difference between the 
full and pointlike fragmentation functions due only to 
the VMD input is small in either scheme, 
and particularly at large $x$.
Furthermore, we see that although the fragmentation function 
in one scheme may be well behaved as 
$x \to 1$, in a different scheme, 
it will diverge as $\ln(1-x)$.   
Surprisingly, it appears well behaved in the $\overline{{\rm MS}}$-scheme 
rather than the specially constructed $DIS_{\gamma}$ scheme. As discussed earlier, this is due to an inaccuracy in the numerical
resummed results produced by inverting only a finite number of moments.
If the large $x$ region is treated correctly, 
the $DIS_{\gamma}$ fragmentation function 
should be well behaved while the $\overline{{\rm MS}}$ fragmentation
function 
will exhibit a logarithmic enhancement.
 
\begin{figure}[t]
\begin{center}
~\epsfig{file=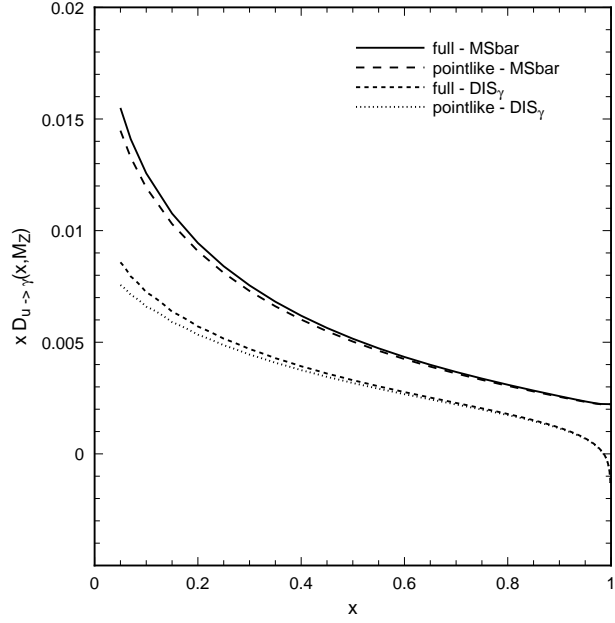, width=9cm} 
\caption{
The GRV up quark-to-photon fragmentation function 
$x D_{u \to \gamma}(x,\mu_F)$  
evaluated $\mu_F = M_Z$.    
The full (pointlike) fragmentation function is shown as a solid 
(dashed) line in the $\overline{{\rm MS}}$-scheme and
as a short-dashed (dotted) line in the $DIS_{\gamma}$-scheme (see text).}
\label{fig:grvDu}
\end{center}
\end{figure}  

In order to be able to implement an expanded form 
for the complete quark-to-photon fragmentation function 
in the BLL and NLO  $(\overline{{\rm MS}})$ expressions 
of the cross section given by 
eqs.~(\ref{eq:sigincnlo}), (\ref{eq:sigbll}) 
we need to know such a form in that scheme.
The sum of the expanded expressions for the 
pointlike and hadronic parts of the fragmentation function 
given in eqs.~(\ref{eq:bllexp}) and (\ref{eq:Dhadexpdef}) in the 
$\overline{{\rm MS}}$-scheme reads,
\begin{equation}
\Dq^{(exp.)}\mid_{\overline{{\rm MS}}}(x,\mu_F)= 
\Dq^{pl,(exp.)}\mid_{\overline{{\rm MS}}}(x,\mu_F)+
\Dq^{had,(exp.)}\mid_{\overline{{\rm MS}}}(x,\mu_F).
\end{equation}
Rewriting the hadronic input in the $DIS_{\gamma}$ scheme yields,
\begin{eqnarray}
\Dq^{(exp.)}\mid_{\overline{{\rm MS}}}(x,\mu_F)  
&=&
\Dq^{pl,(exp.)}\mid_{\overline{{\rm MS}}}(x,\mu_F)\nonumber \\
&-&\aqed\Delta C_{\gamma}^{(0)} \mid_{DIS_{\gamma}}(x)\nonumber \\
&-&\aqed\as\ln \left(\frac{\mu^2_{F}}{\mu_{0}^2}\right)\Pqqzero\otimes 
\Delta C_{\gamma}^{(0)} \mid_{DIS_{\gamma}}(x)\nonumber \\
&+&D_{q \to \gamma}^{np}\mid_{DIS_{\gamma}}(x,\mu_{0})\nonumber \\
&+&
\as\ln \left(\frac{\mu^2_{F}}{\mu_{0}^2}\right)\Pqqzero\otimes 
D_{q \to \gamma}^{np}\mid_{DIS_{\gamma}}(x,\mu_{0}). 
\label{eq:Dnlo3}
\end{eqnarray}
The last two lines of this equation are proportional to the VMD input,
$D_{q \to \gamma}^{np}\mid_{DIS_{\gamma}}(x,\mu_{0})$, which is 
negligible \cite{grv}, and can be neglected.
Note that, an equivalent way to obtain this expression, 
with the model-dependent 
input $D_{q \to \gamma}^{np}\mid_{DIS_{\gamma}}(x,\mu_{0})$ neglected,  
is to consider $\Dq^{np}(x,\mu_{0}) \mid_{\overline{{\rm MS}}}$ 
to be given simply by $-\Delta C_{\gamma}^{(0)} \mid_{DIS_{\gamma}}$.

\subsubsection{Direct contributions evaluated conventionally}
\label{subsubsec:expcon}
The expression for the BLL 
single-photon production cross section in the $\overline{{\rm MS}}$-scheme 
obtained using a conventional power counting
of the direct terms and using the expanded 
expression of the fragmentation function given in eq.~(\ref{eq:Dnlo3}) 
(with $D_{q \to \gamma}^{np}\mid_{DIS_{\gamma}}(x,\mu_{0})=0$) then reads,
\begin{eqnarray}
\frac{1}{\sigma_{0}}\frac {{\rm d} \sigma^{BLL (exp.)}}{{\rm d}x}
&=&\sum_{q=1}^{2N_{f}}\left ( - \aqed \Delta C_{\gamma}^{(0)} \mid_{DIS_{\gamma}}
- \aqed\as\ln \left(\frac{\mu^2_{F}}{\mu_{0}^2}\right)\Pqqzero
 \otimes  \Delta C_{\gamma}^{(0)} \mid_{DIS_{\gamma}} (x)\right .
 \nonumber\\
& &
+ \aqed 
\ln\left(\frac{\mu^2_{F}}{\mu_{0}^2}\right) \Pqpzero(x)
+\aqed\aqed  
\ln \left(\frac{\mu^2_{F}}{\mu_{0}^2}\right)\Pqpone(x)
\nonumber\\
& & +
\frac{1}{2}\aqed \as
\ln^2 \left(\frac{\mu^2_{F}}{\mu_{0}^2}\right) \Pqqzero\otimes \Pqpzero (x)
\nonumber\\
& & \left .+\aqed\as \ln \left(\frac{\mu^2_{F}}{\mu_{0}^2}\right) C_{q}^{(0)}
\otimes \Pqpzero  (x)
+\aqed C^{(0)}_{\gamma}(x,\mu_{F})\right ).
\label{eq:sigbll2}
\end{eqnarray}
where we have only retained terms proportional to 
$\alpha_{s}^{0}$, $(\alpha_{s}^{0}\ln\mu_{F}^2)$, $(\alpha_{s}^{1}\ln\mu_{F}^2)$ and $(\alpha_{s}^{1}\ln^{2}\mu_{F}^2)$. 

\subsubsection{Direct contributions evaluated at fixed order in $\alpha_{s}$}
\label{subsubsec:expfix}
The corresponding  
next-to-leading order expression for the cross section can however 
not be directly obtained by implementing the expanded quark-to-photon fragmentation function of eq.~(\ref{eq:Dnlo3}) 
in the next-to-leading order  
$\overline{{MS}}$ cross section given in eq.~(\ref{eq:sigincnlo}).
We need to consider a further modification to the direct term present in 
eq.~(\ref{eq:sigincnlo}) which is generated 
by the transformation of the non-perturbative input from the
$\overline{{MS}}$-scheme to the $DIS_{\gamma}$-scheme, 
eq.~(\ref{eq:DinputDIS}).
With this change, the term proportional to $C_{q}^{(0)}\otimes \Dq(x,\mu_{F})$
in eq.~(\ref{eq:sigbll}) becomes,
\begin{displaymath}
+\aqed\as C_{q}^{(0)}
\otimes \left[\Pqpzero (x)\ln \left(\frac{\mu^2_{F}}{\mu_{0}^2}\right) + 
\Dq^{np}\mid_{DIS_{\gamma}}(x,\mu_{0})-\Delta C_{\gamma}^{(0)} \mid_{DIS_{\gamma}}\right].
\end{displaymath}
Here, the term proportional to $C_{q}^{(0)}\otimes \Delta C_{\gamma}^{(0)}$
is genuinely of ${\cal O}(\alpha_s)$ and unlike the term proportional to 
$\Pqpzero$ could be ignored in the conventional
BLL approach discussed above.
However, for a consistent treatment at fixed NLO,  it must be retained, 
so that, in the  $\overline{{MS}}$-scheme,
\begin{equation}
\frac{1}{\sigma_{0}}\frac {{\rm d} \sigma^{NLO (exp.)}}{{\rm d}x} =
\frac{1}{\sigma_{0}}\frac {{\rm d} \sigma^{BLL (exp.)}}{{\rm d}x} +
 \aqed \as C_{\gamma}^{(1)} - \aqed \as C_{q}\otimes \Delta C_{\gamma}^{(0)} \mid_{DIS_{\gamma}}.
\label{eq:signlo2}
\end{equation}
Here, we clearly see which terms differ 
between the BLL and NLO cross sections when 
using an expanded expression for the 
fragmentation function.
Another way to understand the origin of this additional term 
is gained by considering the expressions of the next-to-leading order 
cross section ${\rm d}{\sigma}^{NLO}$ 
in both $\overline{{\rm MS}}$ and $DIS_{\gamma}$-schemes.
The requirement that these two quantities are equal,
dictates that,
\begin{displaymath}
\aqed \as C_{\gamma}^{(1)} +\as C_{q}\otimes \Dq(x,\mu_{F})
\end{displaymath}
has to be invariant under the scheme transformation.
Equivalently we have,
\begin{equation}
C_{\gamma}^{(1)}\mid_{DIS_{\gamma}}=C_{\gamma}^{(1)}\mid_{\overline{{\rm MS}}}
-C_{q}\otimes \Delta C_{\gamma}^{(0)} \mid_{DIS_{\gamma}}.
\end{equation}

Let us now turn to the approach followed by BFG to determine 
$\Dq(x,\mu_{F})$.  Unlike the GRV group they consider the evolution  
of the quark-to-photon fragmentation function directly in the 
$\overline{{\rm MS}}$-scheme with a 
non-perturbative input given by,
\begin{equation}
\Dq^{np}\mid_{\overline{{\rm MS}}}(x,\mu_{0})\equiv
\Dq^{VMD}(x,\mu_{0}) - C_{\gamma}^{(col)}(x),
\label{eq:Dnpbfg} 
\end{equation}
where $\Dq^{VMD}$ is fixed by a vector dominance model 
and is negligible.
Effectively, the input is $-C_{\gamma}^{(col)}$,
the collinear part 
of the direct term $C_{\gamma}^{(0)}$ which is 
defined in eq.~(\ref{eq:sigcol}) and again diverges logarithmically
as $x \to 1$.

As mentioned before, 
the GRV group uses the same input and the same evolution equation
for each quark flavour as we do in the fixed order approach.
Unlike in our fixed order approach, 
where all 5 flavours are treated massless, the GRV group considers
the masses of charm and bottom quarks 
($m_{c}=1.5$~GeV and $m_b =4.5~$GeV) 
and let the evolution equations 
of the heavy flavour fragmentation functions  
start at the appropriate mass thresholds \cite{grv}.
In other words, for the charm-to-photon and the bottom-to-photon 
fragmentation function, they select $\mu_{0}=m_{c},m_{b}$ respectively.   
The BFG group also considers the charm and bottom quarks to be massive.
However, in their approach, the input distribution given in 
eq.~(\ref{eq:Dnpbfg}) is valid only for light quarks 
and the heavy quark input is treated slightly differently 
as explained in \cite{bfg}.
For most applications though, 
(such as single photon production at LEP energies)
we are far from 
the quark mass thresholds and the massless evolution of the charm 
and bottom fragmentation functions is a good approximation.
For this reason we will not go into the details of the treatment of 
heavy quarks in the BFG approach
and merely refer the reader to the original work \cite{bfg}.

For reference, we show the full and pointlike 
up-quark fragmentation functions in Fig.~\ref{fig:bfgDu} at $\mu_F = M_Z$.
Here, the difference between the curves is not due to $\Dq^{VMD}$ alone, 
but depends on the combination, $\Dq^{VMD}(x,\mu_{0}) - C_{\gamma}^{(col)}(x)$.
This should engender a significant difference at large $x$, but, 
because of the difficulty of obtaining accurate parameterizations 
at large $x$ numerically, this has been obscured.

\begin{figure}[t]
\begin{center}
~\epsfig{file=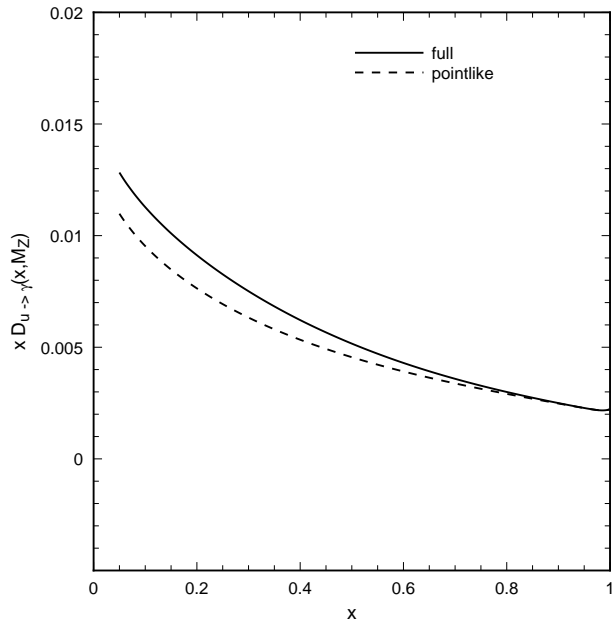, width=9cm} 
\caption{
The BFG up quark-to-photon fragmentation function 
$x D_{u \to \gamma}(x,\mu_F)$  
evaluated at $\mu_F = M_Z$.    
The full (pointlike) fragmentation function is shown as a solid 
(dashed) line in the $\overline{{\rm MS}}$-scheme.}
\label{fig:bfgDu}
\end{center}
\end{figure}  

The analogues of the BLL and NLO expanded expressions  
of the  single photon production cross section following the BFG approach
(at least for the light quark) are obtained by 
replacing $\Delta C_{\gamma}^{(0)} \mid_{DIS_{\gamma}}$
by $C_{\gamma}^{(col)}$ 
in eqs.~(\ref{eq:sigbll2}) and ({\ref{eq:signlo2}). 
A similar transformation accompanied
by replacing $\Dq^{np}(x,\mu_{0})\mid _{DIS_{\gamma}}$  by 
$\Dq^{VMD}(x,\mu_{0})$ in eq.~(\ref{eq:Dnlo3}) yields
an expression for the   
$\overline{{\rm MS}}$ quark-to-photon fragmentation function in the BFG 
approach.
Finally, note that the only difference between the 
expression of the NLO single photon production cross section obtained 
following the fixed order approach with the
BFG non-perturbative input given in eq.~(\ref{eq:Dnpbfg}) and that 
obtained in the fixed order approach as described 
in Section~\ref{sec:ourinclusive} is the different non-perturbative 
input which in our approach is determined by the
ALEPH data, see eq.~(\ref{eq:fitnlo1}).

\subsection{Summary}
\label{subsec:summary}
The different strategies for evaluating the single photon cross section described in this section can be summarized as follows;
\begin{itemize}
\item[I)] Resummed fragmentation function obtained numerically
together with explicit power counting of the coupling constants, 
as described in sec.~\ref{subsubsec:resfix} and at NLO by
eq.~(\ref{eq:sigincnlo})).
\item[II)] Resummed fragmentation function with conventional 
power counting, i.e. associating an inverse power of $\alpha_s$ with $\Dq$.
See sec.~\ref{subsubsec:rescon} and eq.~(\ref{eq:sigbll}) for the BLL result.
\item[III)] Fragmentation function expanded as a series in $\alpha_{s}$
(given in eq.~(\ref{eq:Dnlo3}) and with the non-perturbative input
$D_{q \to \gamma}^{np}\mid_{DIS_{\gamma}}(x,\mu_{0})$ neglected) 
together with conventional power 
counting, i.e. associating an inverse power of $\alpha_s$ with each power 
of $\ln \mu_{F}^2$.
See sec.~\ref{subsubsec:expcon} and eq.~(\ref{eq:sigbll2}).
\item[IV)] Expanded fragmentation function 
together with explicit power counting of the coupling constants, 
as described in sec.~\ref{subsubsec:expfix} and eq.~(\ref{eq:signlo2}).
\end{itemize}

Provided the resummed solution of the all order evolution equation 
can be accurately determined, the approach using this solution 
and the direct terms evaluated at fixed order (approach I) represents 
the theoretically preferred approach.  
The approach evaluating the direct terms at fixed order and 
using an expanded and thereby approximate expression 
of the fragmentation function has however 
important advantages. It enables an analytic determination of the 
fragmentation function and yields factorization scale independent results 
for the photon production cross section evaluated at a given order 
in $\alpha_{s}$. Furthermore, the implementation of 
the non-perturbative quark-to-photon fragmentation function which was 
fitted to the  ALEPH `photon' +~1 jet in the fixed order expressions of the 
`isolated' and inclusive rate yielded perturbative stable 
predictions~\cite{letter} which agreed well with the ALEPH and OPAL data. 

In the next section we shall see how the theoretical predictions obtained 
following any of these approaches and 
using the GRV or BFG schemes compare with the OPAL and ALEPH data.
For definiteness, in approaches I and II we will use either the
parameterization of the pointlike fragmentation function of the GRV group 
or the sum of pointlike and hadronic parts (set I)   
of the BFG group.
In approaches III and IV, we will consider eq.~(\ref{eq:Dnlo3}) 
with the VMD input $D_{q \to \gamma}^{np}\mid_{DIS_{\gamma}}(x,\mu_{0})$ 
set to zero to describe the evolution equations 
for all active flavours in the GRV case, and the same equation with 
$\Delta C_{\gamma}^{(0)} \mid_{DIS_{\gamma}}$
replaced by $C_{\gamma}^{(col)}$ in the BFG case.
As a result, in these approaches and in either of the two schemes 
all quark flavours satisfy a massless evolution equation.  
Finally, the light flavours start their evolution 
at $\mu_{0}=0.55$ GeV and at $\mu_{0}=0.71$ GeV
respectively in the GRV or in the BFG schemes, 
while the heavy $c$ and $b$ quark fragmentation functions
start to evolve at $\mu_0 = m_{c}$ and $\mu_0 = m_{b}$ respectively.

\section{Results}
\setcounter{equation}{0}
\label{sec:results}

In the previous sections we have completed a detailed comparison of 
the two fundamentally different approaches (fixed order and conventional)
for computing photon cross sections.
We have described how both the cross sections 
and fragmentation functions are defined in these two formalisms and
in Section~\ref{sec:comparison} we have given expressions for 
the cross section obtained using an expanded 
expression for the quark-to-photon fragmentation function $\Dq(x, \mu_{F})$.
We have now collected all necessary ingredients 
to be able to evaluate the single photon production cross section 
in any of the four approaches summarized in Section~\ref{subsec:summary}
and could perform this task while defining the non-perturbative input
according to either of the GRV or BFG groups. 
We shall apply these calculations to compute both the
inclusive cross section and the `photon' +~1 jet rate
and make comparisons with the OPAL 
and ALEPH data in the second part of this section.

Before turning to the cross sections,  
we first compare the analytic expanded expression  
of the quark-to-photon fragmentation function given by eq.~(\ref{eq:Dnlo3})
with the numerically resummed BLL results for both GRV and BFG prescriptions.
This is shown in Fig.~\ref{fig:Dq} for the up-quark.   
Note that in addition to fixing the non-VMD input, 
$\Delta C_{\gamma}^{(0)} \mid_{S}$ 
differently in the two schemes,
the hadronic scale is also different, $\mu_0 = 0.55$~GeV for BLL GRV
and $\mu_0 = 0.71$~GeV for BLL BFG.
There are two ranges of interest, $0.2 < x < 0.95 $ which is relevant for the 
inclusive photon data from OPAL and $0.7 < x < 1$ appropriate for the ALEPH
`photon' + 1~jet data.

We see that, except in the very high $x$ region, the various fragmentation functions generally agree well with each other in shape and magnitude.
As discussed earlier,
at large $x$, there are significant disagreements which are mainly due to
deficiencies in the numerical parameterizations\footnote{In fact,
it is striking to notice that the only quark-to-photon fragmentation 
functions which appear to diverge as $x\to 1$ are those 
which have at least one analytic component which itself diverges as $x \to 1$.}.
We therefore expect, that predictions for the inclusive photon cross sections
(which run over a wide range of $x$) will be largely in agreement, while
significant differences may be apparent in the `photon' + 1~jet estimates which focus on the large $x$ region.

We also see that the expanded fragmentation functions 
defined according to the BFG and GRV prescriptions are quite different.   This is in part due to the different choice of hadronic scale, but mainly due to the fact that the non-VMD BFG input is more negative than that for GRV.   As can be seen from eq.~(\ref{eq:Dnlo3}) and the definitions of $\Delta C_{\gamma}^{(0)} \mid_{DIS_{\gamma}}$ and
$C_{\gamma}^{(col)}$, then
for the same hadronic scale,
\begin{equation}
\Dq^{(exp.)GRV}\mid_{\overline{{\rm MS}}}(x,\mu_F)-
\Dq^{(exp.)BFG}\mid_{\overline{{\rm MS}}}(x,\mu_F)
> 0.
\end{equation}

\begin{figure}[t]
\begin{center}
~ \epsfig{file=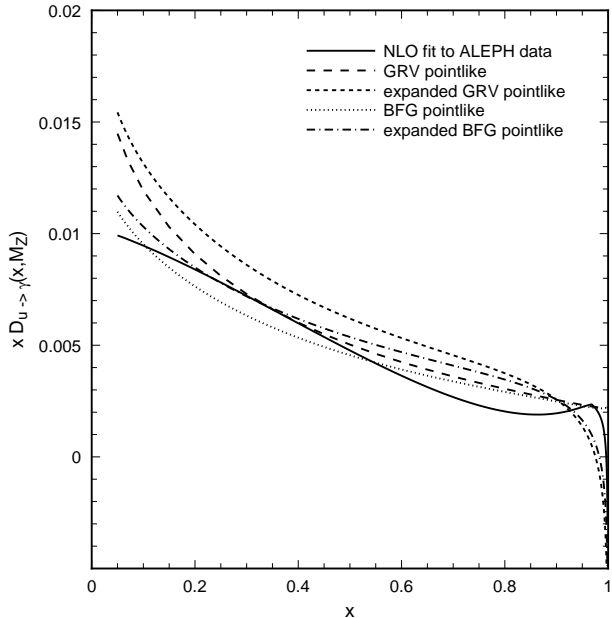, width=9cm} 
\caption{The quark-to-photon fragmentation function 
$x D_{u \to \gamma}(x,\mu_F)$  
evaluated at $\mu_F = M_Z$ in the $(\overline{{\rm MS}})$-scheme. 
The NLO fit from the ALEPH `photon' + 1~jet data is shown
as solid line.   The pointlike GRV (BFG) parameterization is shown
dashed (dotted) while the expanded result of eq.~(\ref{eq:Dnlo3}) is
shown short-dashed (dot-dashed).}
\label{fig:Dq}
\end{center}
\end{figure}  

\subsection{The inclusive cross sections}

In this subsection we collect the results obtained 
evaluating the inclusive one-photon production
cross sections following any of the four approaches described in Section 4.

So far, we have ignored the gluon-to-photon fragmentation function throughout.
To illustrate the tiny role the gluon-to-photon fragmentation plays in
a physical cross section, 
Fig.~\ref{fig:gluon} shows the BLL prediction for the inclusive rate
for the pointlike GRV
parameterization both with and without the gluon-to-photon fragmentation 
contribution.
To make the small difference manifest, we have multiplied the 
gluon-to-photon fragmentation contribution by a factor of 100.
We see that at large $E_{\gamma}$, even when multiplied by a factor of 100, 
the gluon fragmentation contribution is entirely negligible.
At lower energies, the gluon fragmentation reduces the cross section
by at most 5\% at $E_{\gamma}\sim 10$~GeV.
Similar results are also obtained at LL. 
In the following we shall therefore ignore the gluon-to-photon 
fragmentation contribution to the photon production cross section.

\begin{figure}[t]
\begin{center}
~ \epsfig{file=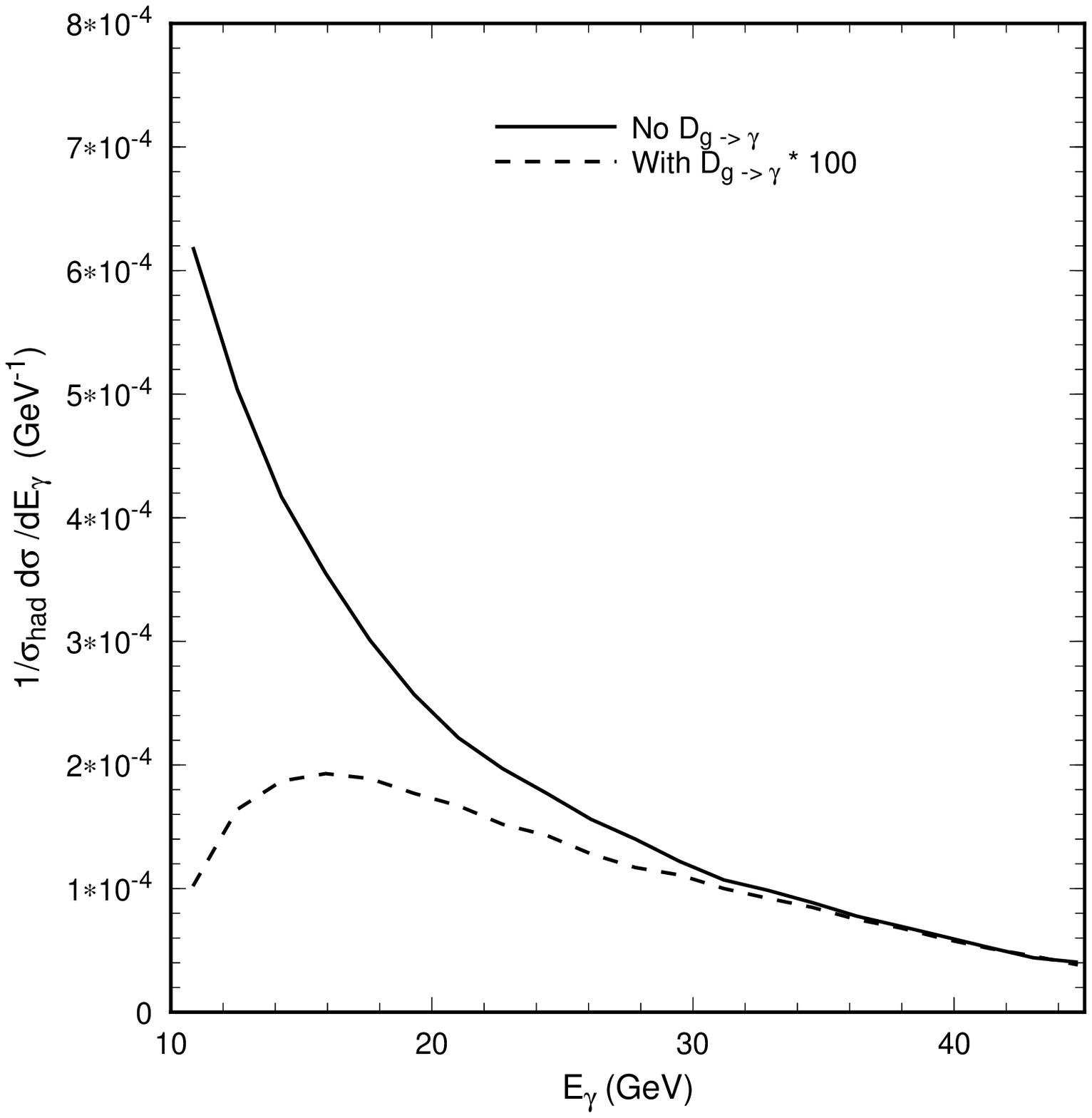, width=9cm} 
\caption{The BLL prediction for the inclusive rate evaluated at $\mu_F = M_Z$ 
using the pointlike GRV parameterization with the gluon-to-photon 
fragmentation function 
ignored (solid) and multiplied by 100 (dashed).
The experimental data is taken from \cite{opal}.}
\label{fig:gluon}
\end{center}
\end{figure}
  
\subsubsection{Perturbative stability and $\mu_{F}$-dependence}

The results obtained using the GRV parameterization
together with conventional power counting at LL and BLL (i.e. approach II)
are shown in Fig.~\ref{fig:incscaleII}.
The upper/lower curves correspond to varying the factorization scale
in the range $2M_Z/0.5M_Z$.
We see that the LL and BLL predictions are similar and thus appear to be 
perturbatively stable.
Furthermore, the factorization scale dependence is significantly
reduced in going from LL to BLL.

\begin{figure}[t]
\begin{center}
~ \epsfig{file=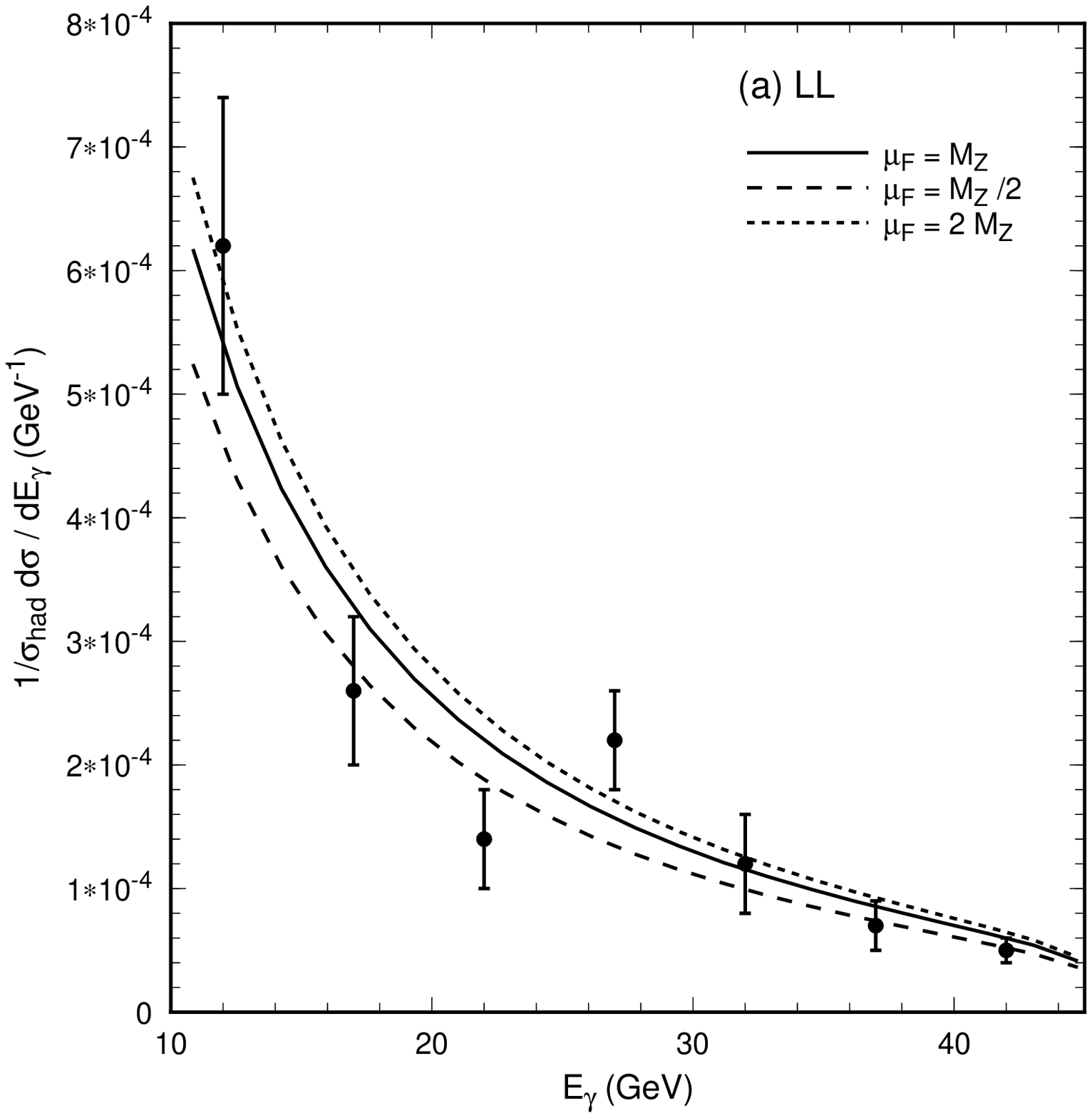, width=7cm} 
~ \epsfig{file=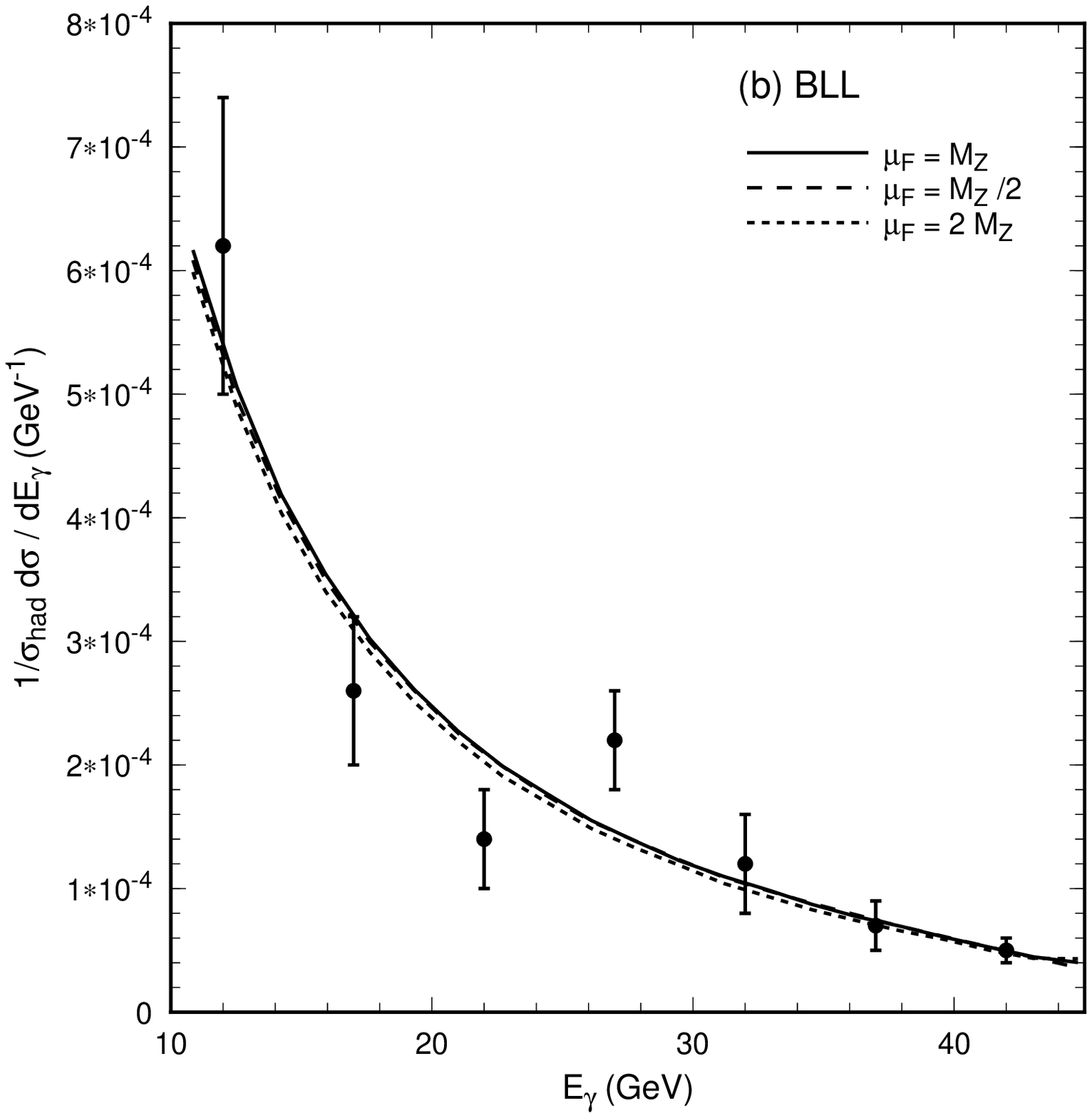, width=7cm} 
\caption{The (a) LL and (b) BLL predictions for the inclusive photon rate
using the pointlike GRV parameterization and conventional power counting.
The solid lines show the prediction for 
$\mu_F = M_Z$, while the short-dashed (long-dashed) lines 
show the expected rate for $\mu_F = 2 M_Z~(0.5 M_Z$).
The experimental data is taken from \cite{opal}.}
\label{fig:incscaleII}
\end{center}
\end{figure}  

Let us see what happens when the inclusive rate 
is evaluated in approach I, 
i.e. using the resummed LL and BLL fragmentation function 
but with the direct contributions evaluated at 
fixed order. This is shown in Fig.~\ref{fig:incscaleI} at LO and NLO 
for the same three choices of the factorization scale 
as in Fig.~\ref{fig:incscaleII}.
In this case, it appears that we can draw the same conclusion regarding the 
$\mu_{F}$ dependence, it is reduced in going from LO to NLO.
However, the LO and NLO results 
appear to be significantly different over the  
whole range of $E_{\gamma}$. This difference is caused by 
the presence of the direct term $C_{\gamma}^{(0)}$ in the LO and NLO 
expressions of the cross section (see eq.~(\ref{eq:siginclo}) and 
(\ref{eq:sigincnlo})).
The failure to describe the inclusive data with  
the LL resummed fragmentation function and with
the direct contributions evaluated at lowest order 
in the expression of the cross section indicate that 
the direct and fragmentation contributions
are not properly matched to each other when the cross section is evaluated 
in this way.

\begin{figure}[t]
\begin{center}
~ \epsfig{file=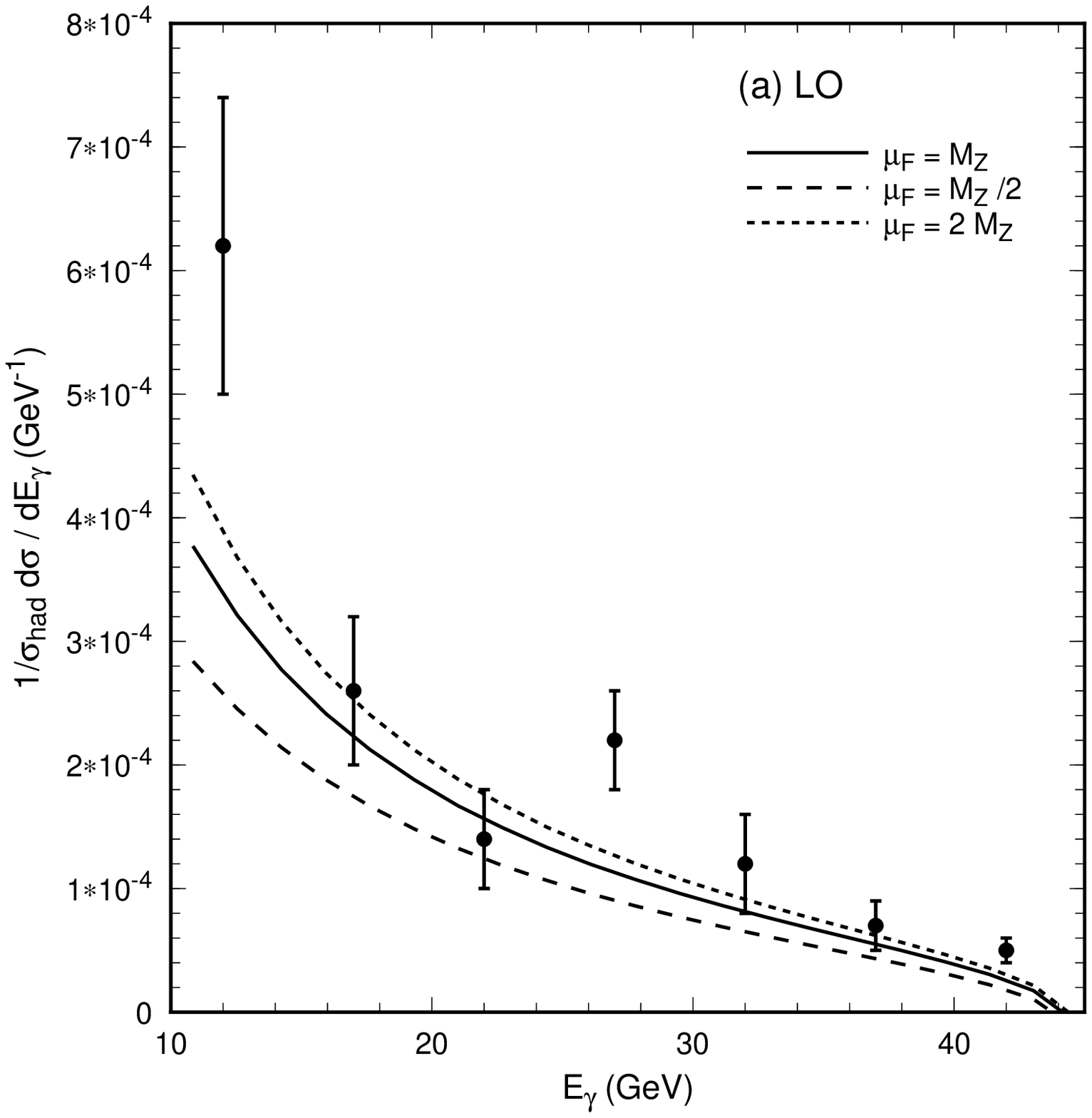, width=7cm} 
~ \epsfig{file=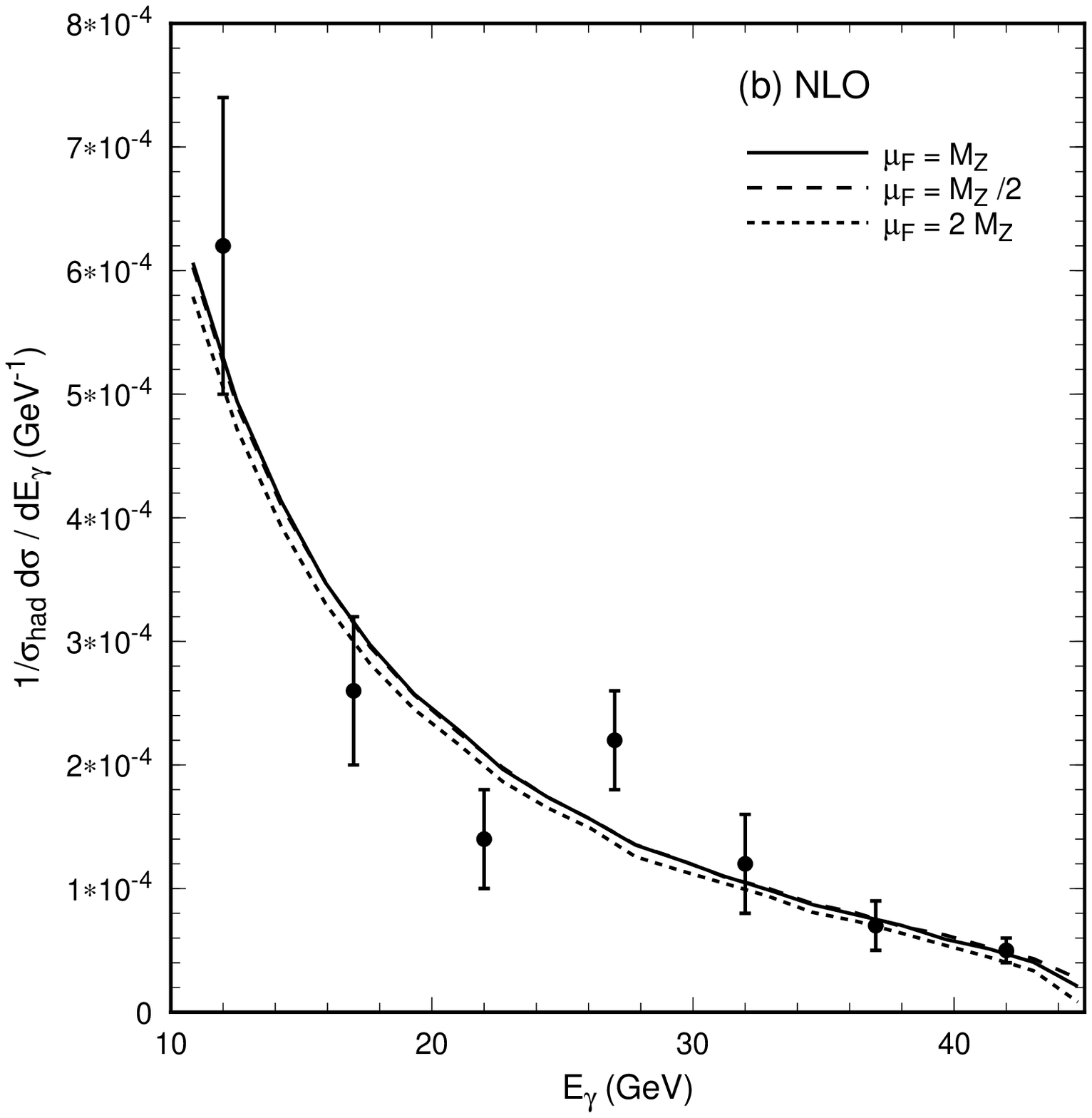, width=7cm} 
\caption{The (a) LO and (b) NLO predictions for the inclusive photon rate
using the pointlike GRV parameterization and explicit power counting.
The solid lines show the prediction for 
$\mu_F = M_Z$, while the short-dashed (long-dashed) 
lines show the expected rate for $\mu_F = 2 M_Z~(0.5 M_Z$).
The experimental data is taken from \cite{opal}.}
\label{fig:incscaleI}
\end{center}
\end{figure}  

\subsubsection{Comparison of results}

\begin{figure}[t]
\begin{center}
~ \epsfig{file=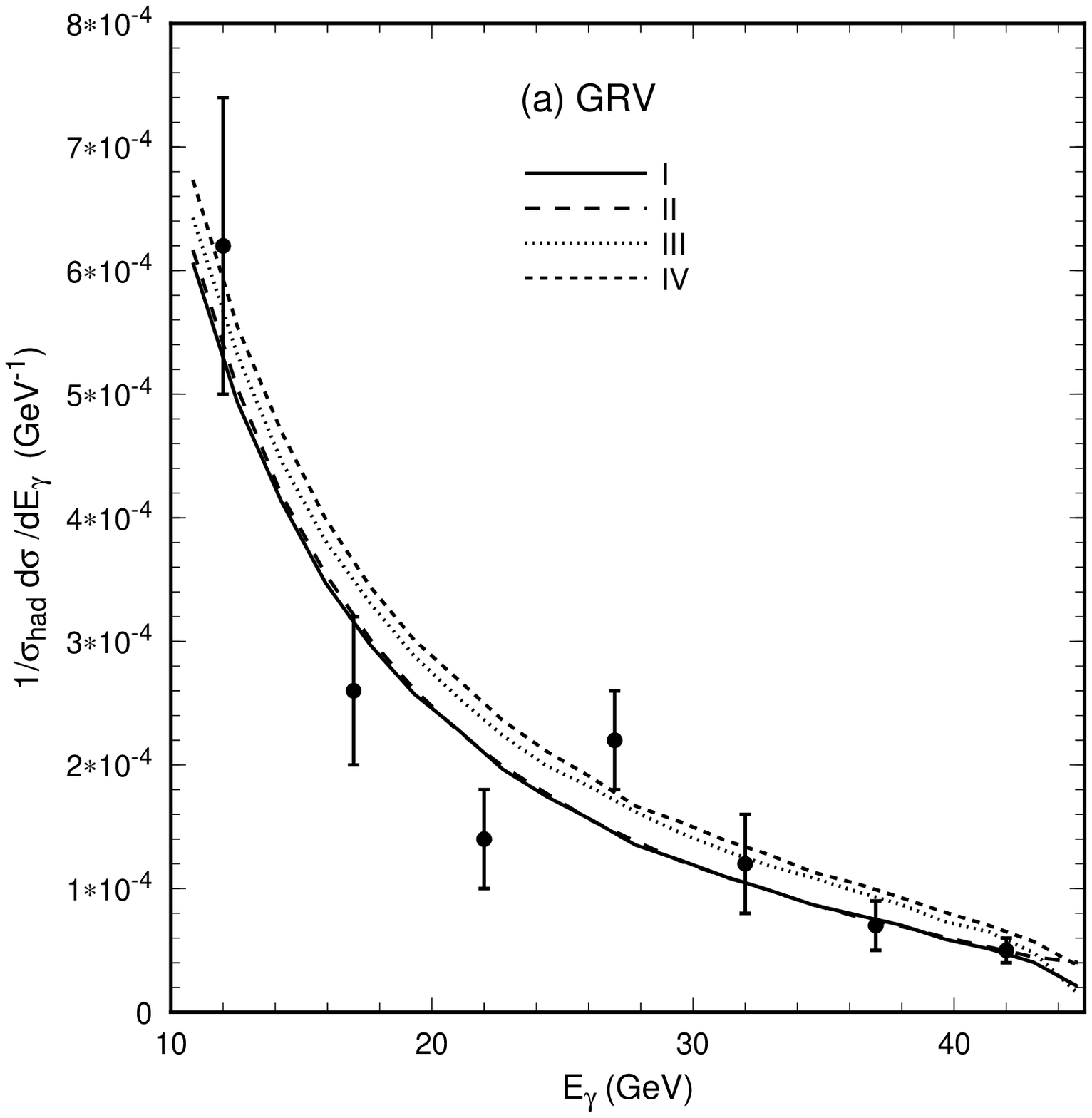, width=7cm}
~ \epsfig{file=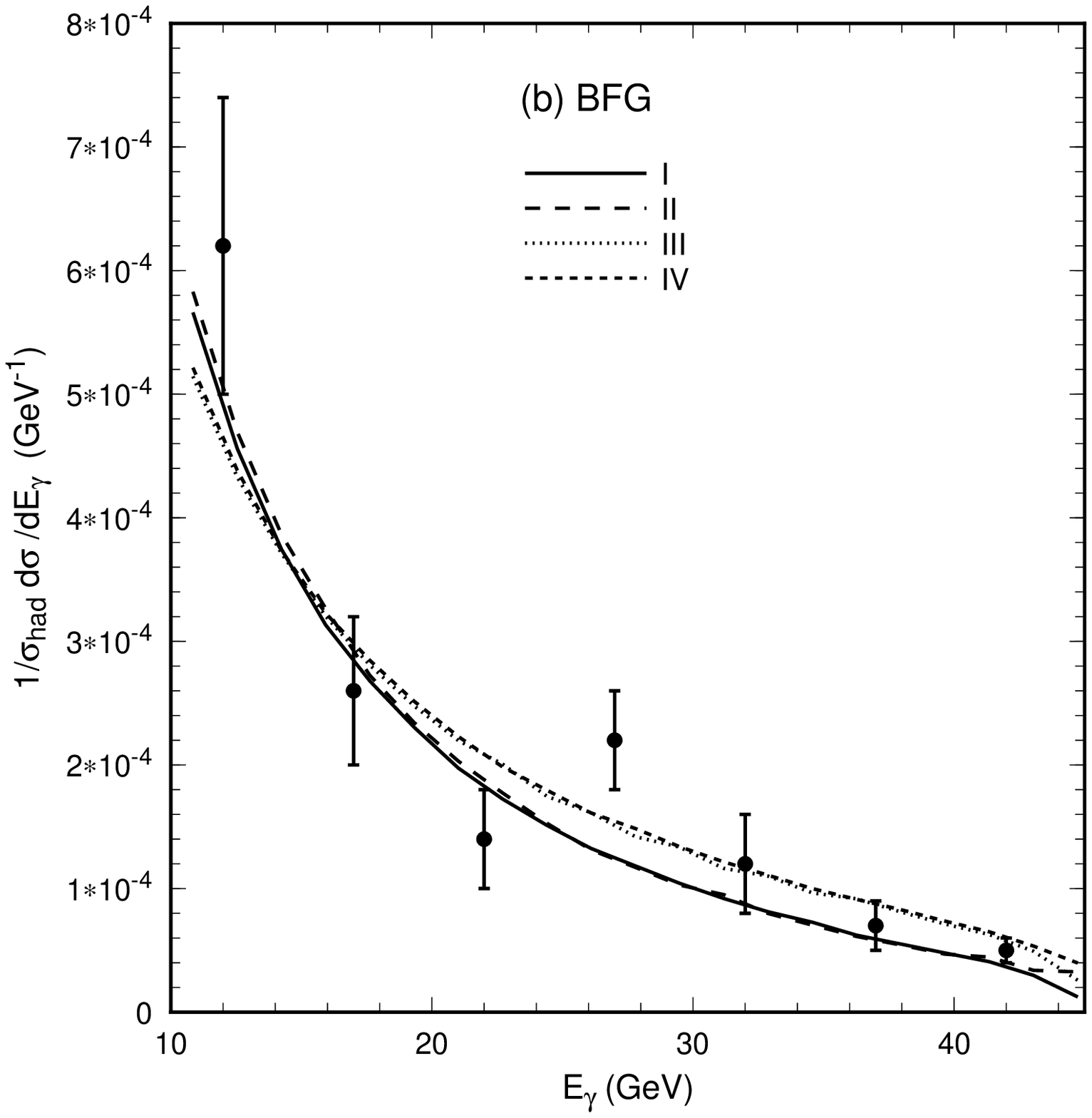, width=7cm} 
\caption{The inclusive photon spectrum estimated from the 
four approaches discussed in sec.~\ref{sec:comparison}
using (a) the GRV and (b) the BFG parameterizations.
The experimental data is taken from \cite{opal}.}
\label{fig:incfinal}
\end{center}
\end{figure} 

Fig.~\ref{fig:incfinal} shows the (NLO or BLL)
inclusive cross section obtained 
using each of the four approaches described in Section~\ref{sec:comparison}
for $\mu_{F}=M_{Z}$.
We show the predictions using both the GRV and BFG schemes while considering  
in each case the definitions of the fragmentation function given in Section 4. 
The various approaches give predictions which
have a similar shape and lie in a common band which is well 
contained within the experimental 
error bars over the whole $x$ range of the
OPAL data. 
The agreement between the predictions is
largely due to the similarity between the 
fragmentation functions as shown in Fig.~\ref{fig:Dq}
but also because the leading direct term $C_{\gamma}^{(0)}$ 
is included in each approach.
We note that the cross section obtained using the expanded 
expression for the fragmentation function (III and IV)
lies by an almost constant amount  
above the prediction obtained using the resummed 
fragmentation functions (I and II)
over the whole $x$ range of interest.
However, given the size of the experimental errors,
all four predictions appear to describe the OPAL data equally well. 

\subsection{The `photon' +~1 jet rates}

Let us now present the results obtained for the `photon' +~1 jet rate.
In the following, we focus on  
one particular value of the jet clustering parameter $\ycut$,
$\ycut=0.1$.
We note that in the range of interest for 
the ALEPH data, $0.7<z<1$ 
the difference between the results obtained using the pointlike 
quark-to-photon fragmentation functions 
alone or the full fragmentation function (i.e. sum 
of pointlike and and hadronic parts) is small. Furthermore, 
the gluon-to-photon fragmentation plays an entirely negligible role.

\subsubsection{Perturbative stability and $\mu_{F}$ dependence} 

\begin{figure}[t]
\begin{center}
~ \epsfig{file=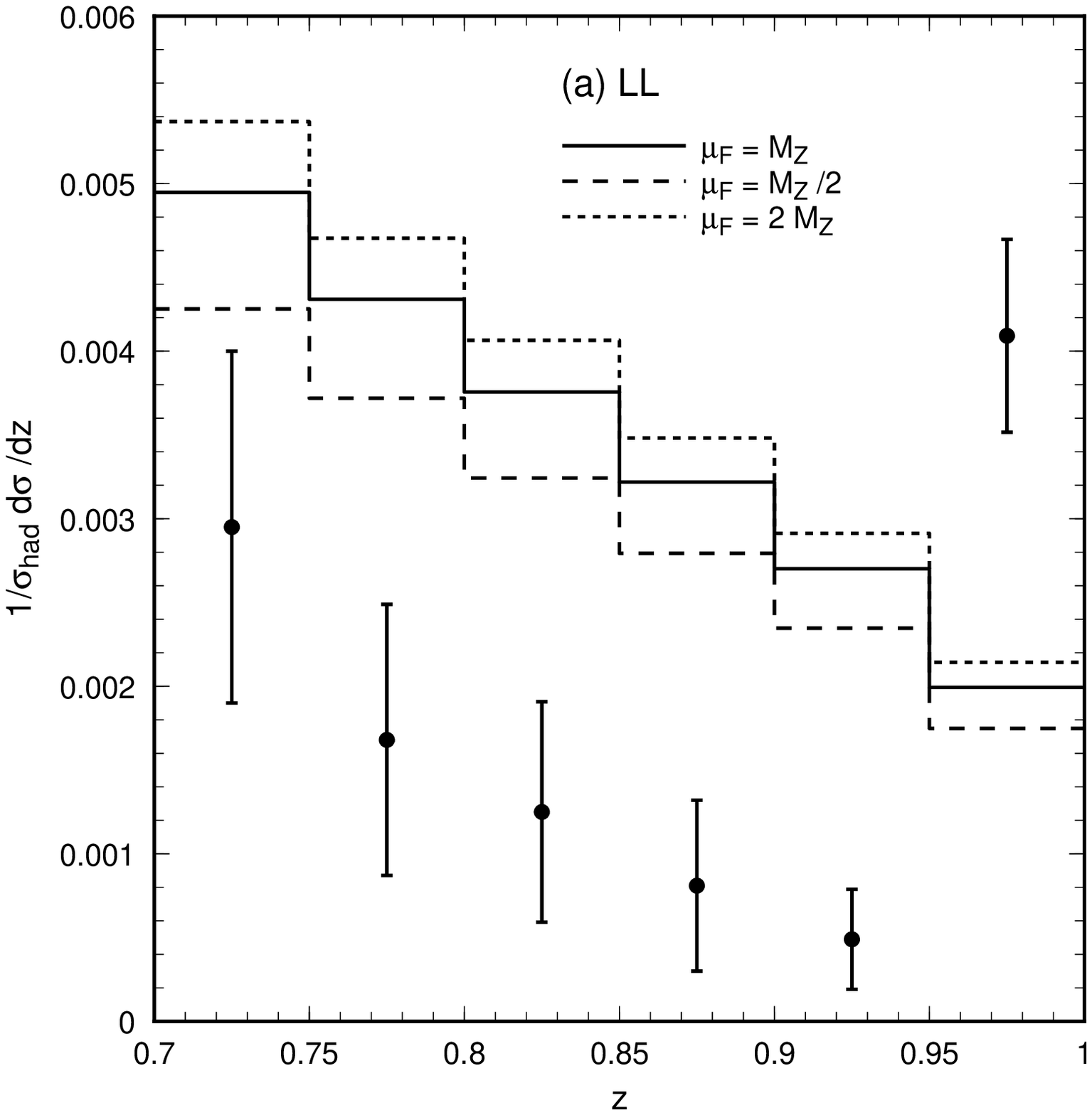, width=7cm} 
~ \epsfig{file=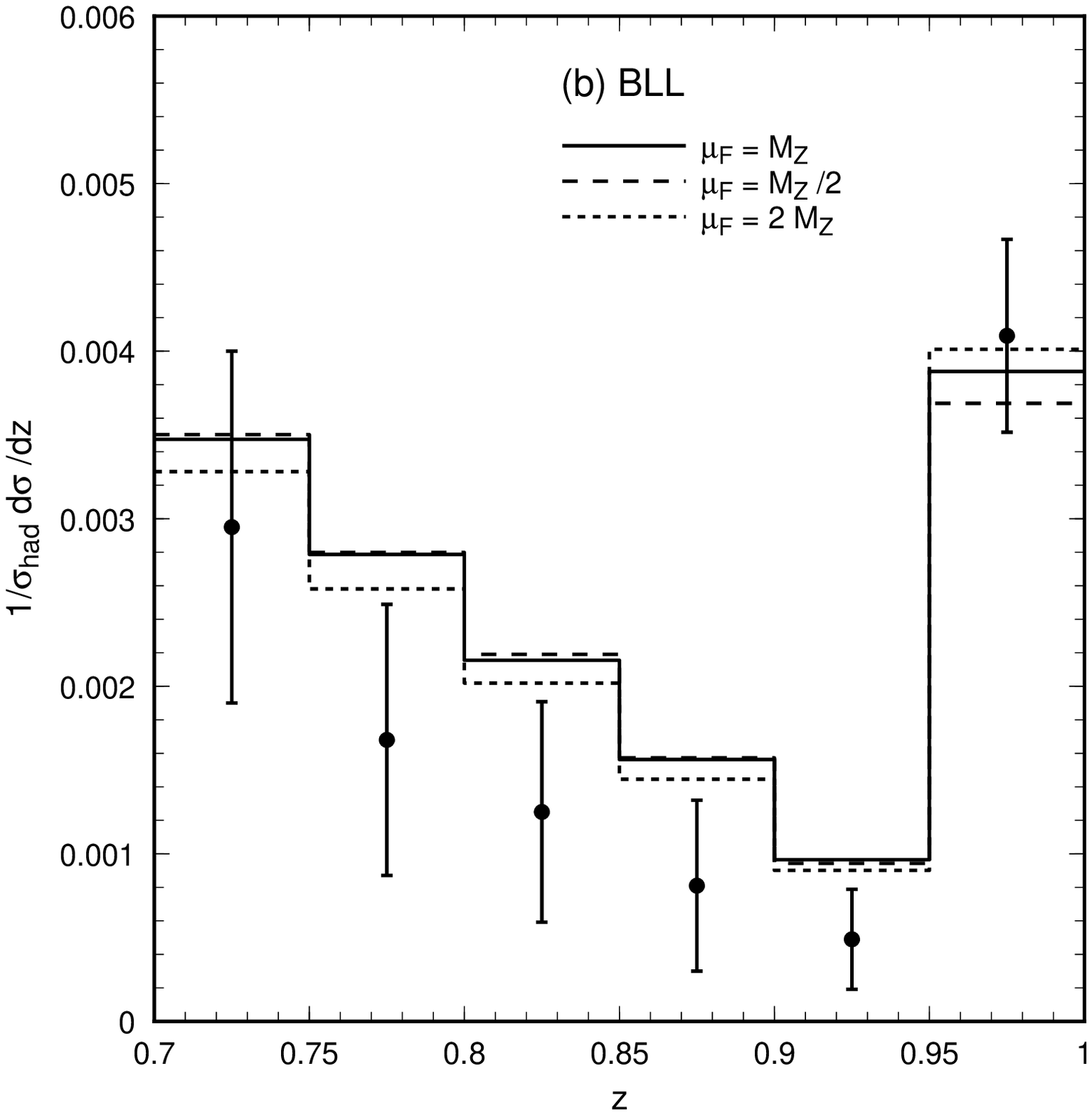, width=7cm} 
\caption{The (a) LL and (b) BLL predictions for the `photon' + 1 jet rate
using the pointlike GRV parameterization and conventional power counting.
The solid lines show the prediction for $\mu_F = M_Z$, while the short-dashed 
(long-dashed) lines show the expected rate for $\mu_F = 2 M_Z~(0.5 M_Z$).
The experimental data is taken from \cite{aleph}.}
\label{fig:jetscaleII}
\end{center}
\end{figure}

Predictions for the `photon' + 1~jet rate
using the pointlike GRV parameterization
together with conventional power counting at LL and BLL (i.e. approach II)
are shown in Fig.~\ref{fig:jetscaleII}.
We vary the factorization scale over a factor of 2 of
the central scale $\mu_F = M_Z$, and, as before, 
the factorization scale dependence is significantly
reduced in going from LL to BLL.
However, the difference between the LL and BLL results is sizeable 
and the shape completely different. In particular, the BLL prediction does 
match the shape of the data quite well.

\begin{figure}[t]
\begin{center}
~ \epsfig{file=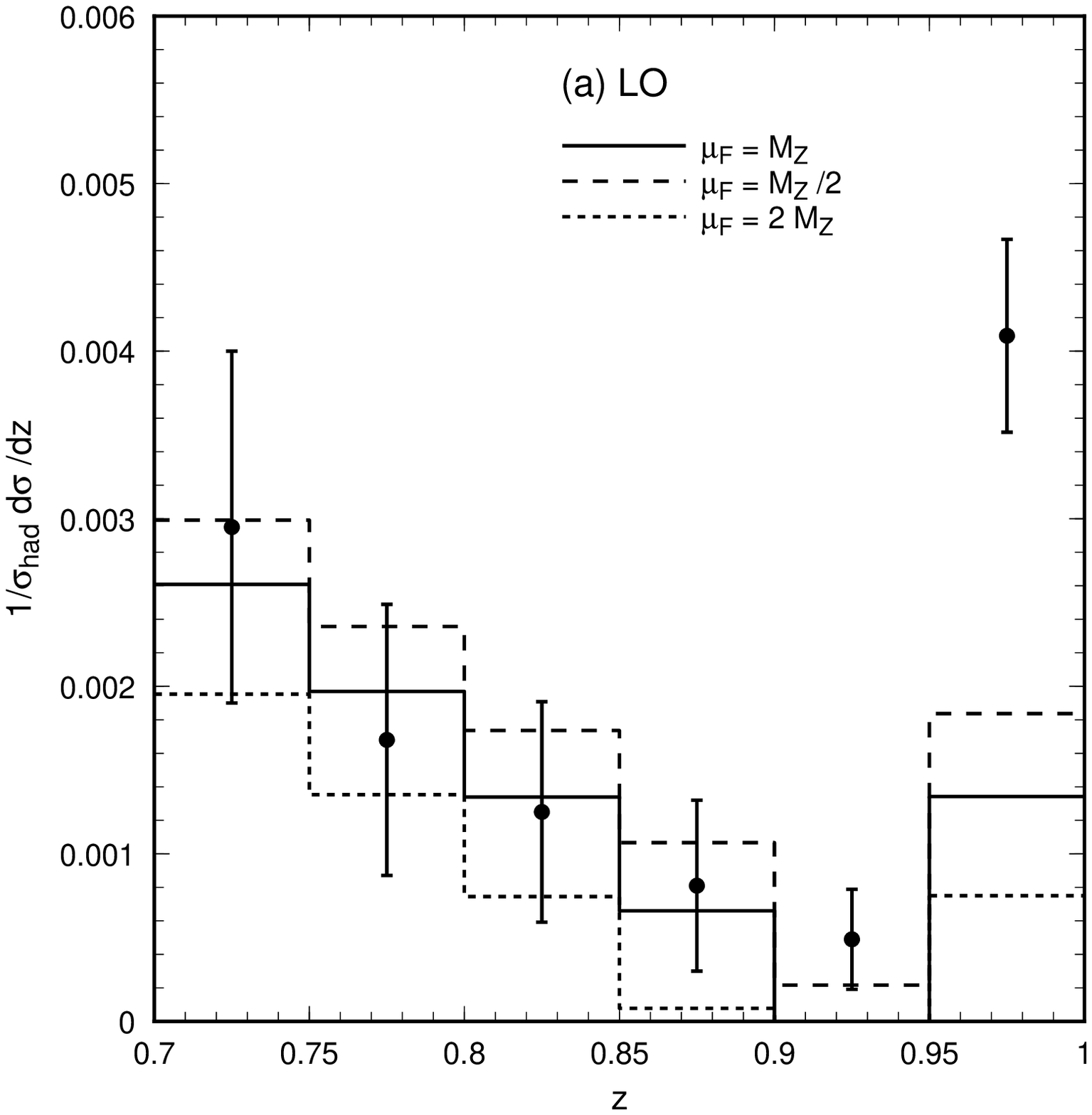, width=7cm} 
~ \epsfig{file=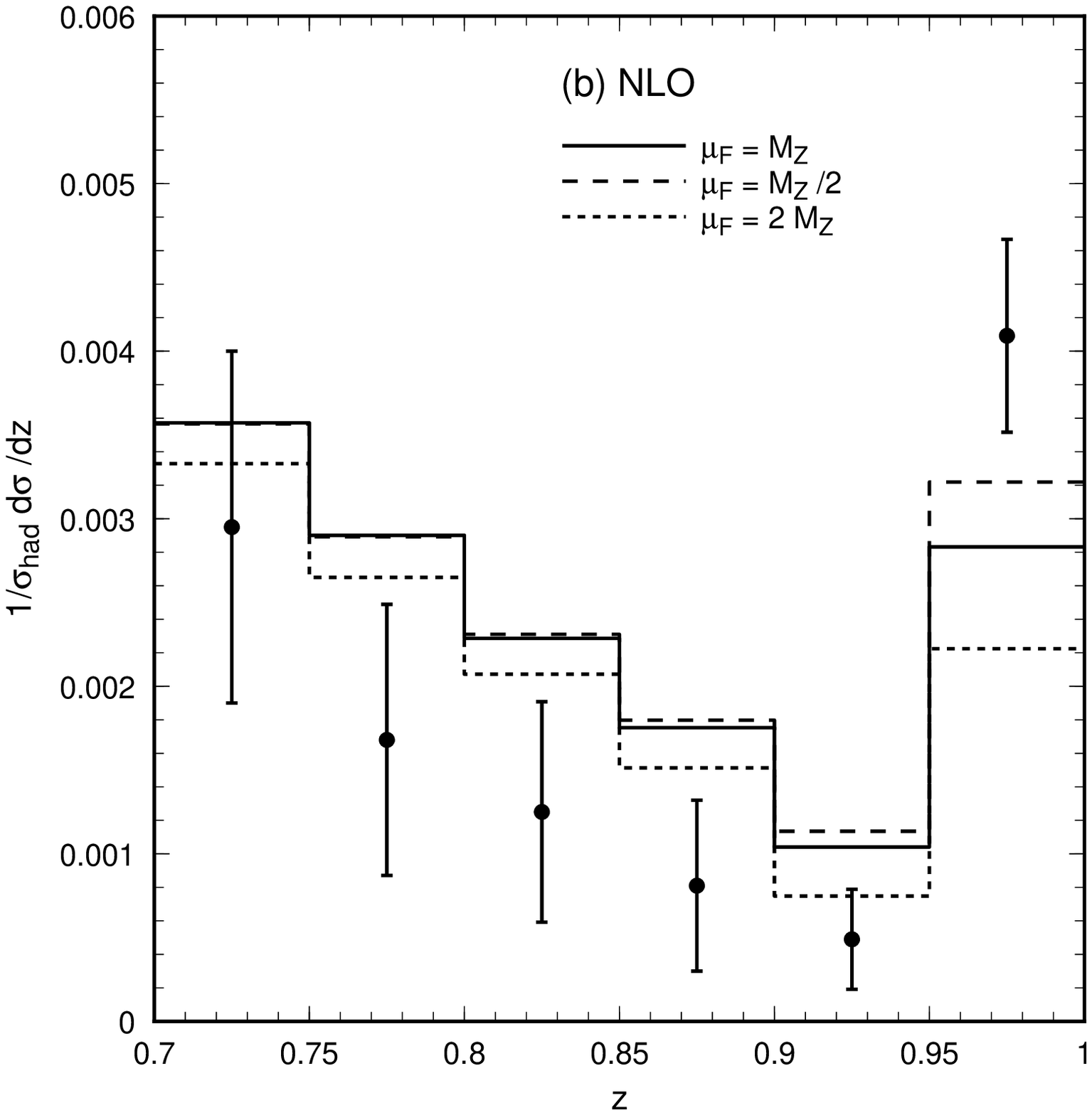, width=7cm} 
\caption{The (a) LO and (b) NLO predictions for the `photon' + 1 jet rate
using the pointlike GRV parameterization and conventional power counting.
The solid lines show the prediction for $\mu_F = M_Z$, while the short-dashed 
(long-dashed) lines show the expected rate for $\mu_F = 2 M_Z~(0.5 M_Z$).
The experimental data is taken from \cite{aleph}.}
\label{fig:jetscaleI}
\end{center}
\end{figure}  

Let us now analyze what happens if one considers the resummed expression 
for the GRV fragmentation function 
in an expression of the cross section where the direct terms 
are evaluated at fixed order in $\alpha_{s}$ (i.e. approach I).
Fig.~\ref{fig:jetscaleI} shows the LO and NLO
`photon' + 1 jet predictions for the same three values 
of the factorization scale $\mu_{F}$.
As can be seen by comparing the leading and next-to-leading order results,
the factorization scale dependence is significantly reduced. 
We also see that
the shapes of the curves displayed at leading and next-to-leading order 
are not dramatically changed. This could be viewed as an indication 
that the results obtained in this approach are perturbatively stable.
For $\mu_{F}=M_{Z}$ we find that, 
the lowest order prediction appears to be slightly below the data 
while the next-to-leading order result lies above the data.
 
\subsubsection{Comparison of results}

\begin{figure}[t]
\begin{center}
~ \epsfig{file=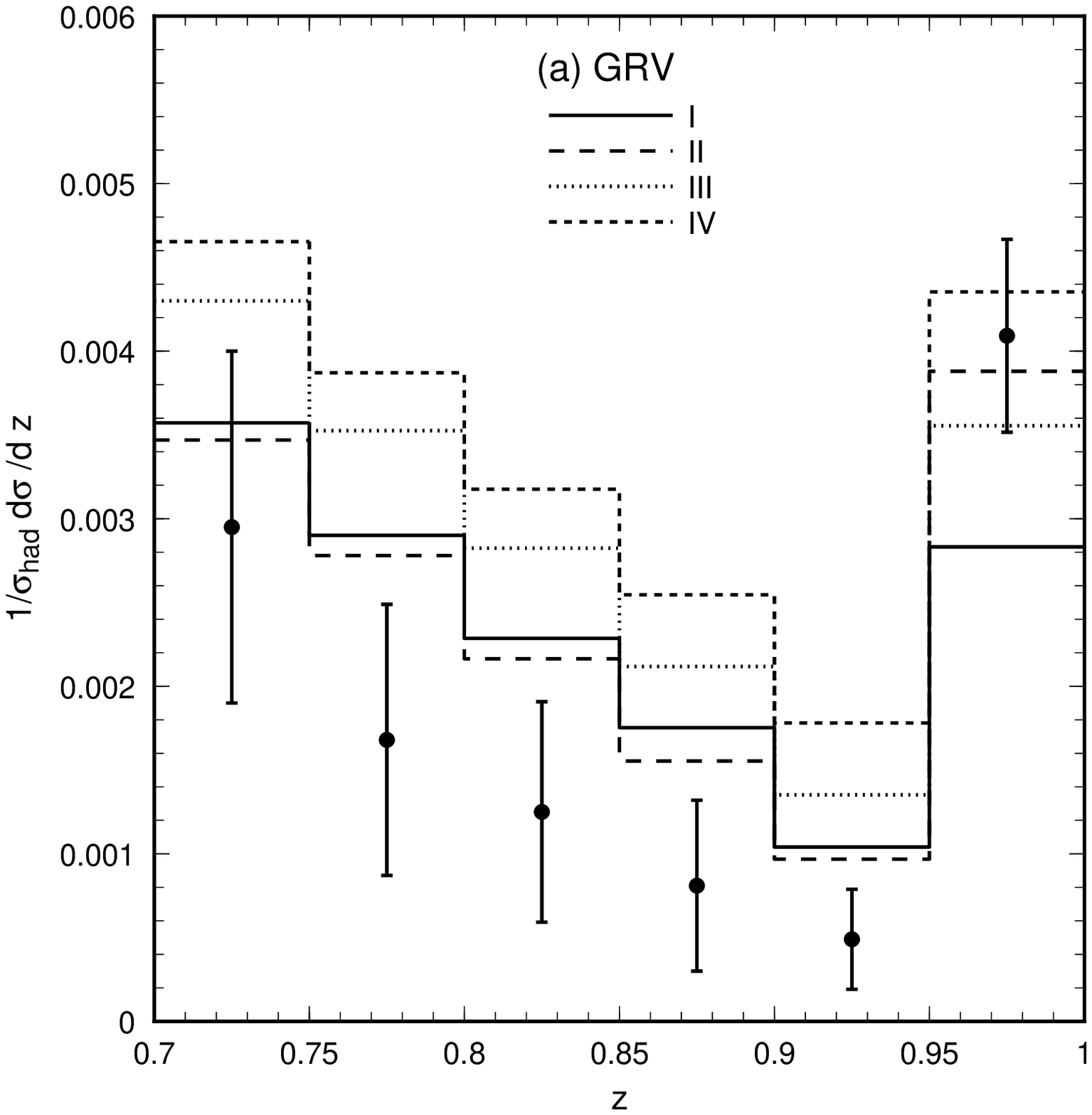, width=7cm}
~ \epsfig{file=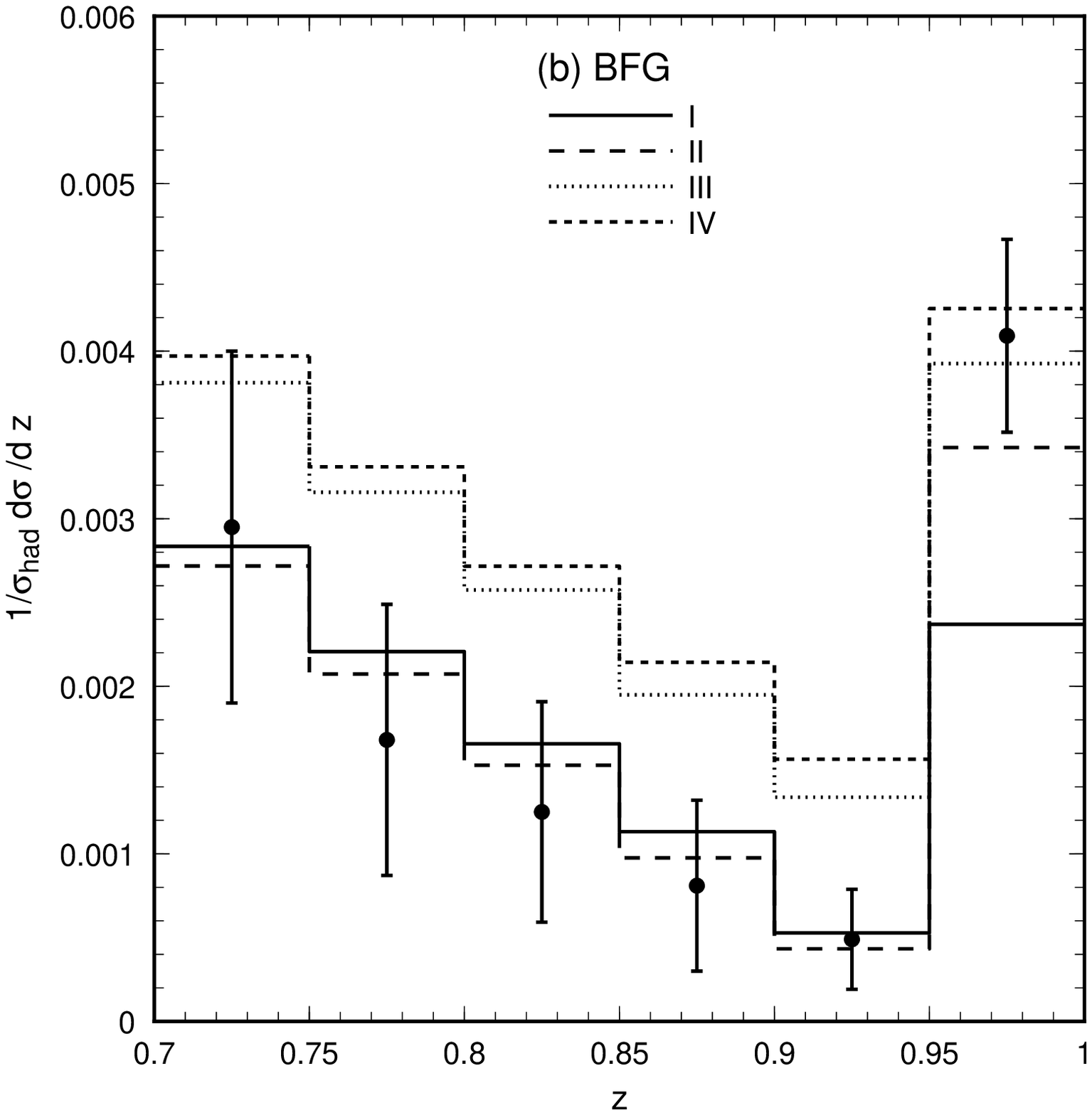, width=7cm} 
\caption{The `photon' + 1~jet rate estimated from the 
four approaches discussed in sec.~\ref{sec:comparison}
using (a) the GRV (b) the BFG inputs.
The experimental data is taken from \cite{aleph}.}
\label{fig:jetfinal}
\end{center}
\end{figure}  

Fig.~\ref{fig:jetfinal} shows the (NLO or BLL)
`photon' + 1~jet rates  obtained 
using each of the four approaches described in Section~\ref{sec:comparison}
for $\mu_{F}=M_{Z}$, with the fragmentation 
functions used in each approach as defined in Section 4.
Ignoring the large $z$ region where we have reason to
doubt the accuracy of the
parameterizations in methods I and II,
we see that the BFG predictions lie systematically below that obtained using 
the GRV parameterization and go through the experimental data points.
As discussed earlier, 
this difference is due to both the choice of hadronic scale and 
the non-VMD input.  The BFG input is smaller and the `photon' + 1~jet
data clearly selects this choice.
Notice however, that the BFG parameterization for the fragmentation function 
unlike that of the GRV group was proposed well after the ALEPH data were released.
As in the inclusive photon rate, predictions involving the expanded
fragmentation function (approaches III and IV) always 
lie above the corresponding approach using the resummed fragmentation
function (I and II).  Again, the data clearly prefers the resummed
fragmentation function. However, the shape of the predictions obtained with
an expanded fragmentation function indicates that adding a negative constant
to them would describe the data very well.

\section{Conclusions}
\setcounter{equation}{0}
\label{sec:conclusions}

In this paper we have made a detailed study of photon production
in hadronic events in electron-positron annihilation at LEP energies.
First, we have used the fixed order approach of \cite{andrew,letter,paper}
to estimate the inclusive photon spectrum and to compare it with the recent
OPAL data \cite{opal}.
Here, the fragmentation function is determined at large $x$ ($x > 0.7$) 
by the ALEPH `photon' +~1 jet data \cite{aleph} and 
is an exact solution of the evolution equation 
without resummation of logarithms of the factorization scale.
As such, the prediction is scale independent and,
surprisingly, agrees well
with the OPAL data which corresponds to $x$ values 
\footnote{Note that most of the events would be categorized as 
`photon' + ~1 jet events 
if a jet algorithm had been applied 
(See Fig.~\ref{fig:plot1}).}  as small as 0.2.  
This is a powerful indication that the fragmentation function fitted to 
the ALEPH data is process independent and can be used to predict photon
cross sections in other processes. 

Alternative methods to compute inclusive photon cross sections
rely on numerically solving the 
evolution equations with some non-perturbative input.  
This input has two pieces, a small vector meson dominance contribution 
together with a perturbative counterterm.   Different parameterizations 
deal with this ambiguity in different ways.  We have examined the 
choices made by the BFG and GRV groups and used them to compute both the 
inclusive photon and `photon' + 1~jet rates. 
To check the general behaviour of the fragmentation function, we have 
made an analytic series expansion in the strong coupling.   
As a result, we find that the large $x$ behaviour of the 
fragmentation functions is not well reproduced by the 
parameterizations, the main problem being to describe 
a logarithmic behaviour with a polynomial.

An additional subtlety is that although the fragmentation function 
appears to be ${\cal O}(\alpha)$, inspection of the evolution equation suggests
a logarithmic growth with $\mu_F$, and in many analyzes, it is ascribed 
a nominal power of $\alpha/\alpha_s$.  Constructing the cross section at some
particular order depends on this assignment 
and different terms will contribute.
This was discussed at length in Section~\ref{sec:comparison}.
In addition, the gluon-to-photon fragmentation function is naively
of ${\cal O}(\alpha\alpha_s)$ and is expected to be much smaller than the
quark contribution.   This is indeed the case for physical cross sections in 
electron-positron annihilation where gluon production is suppressed,
and we ignore the gluon fragmentation function contribution throughout.

In order to better isolate the differences between the expressions of the 
cross section evaluated in a fixed order or in a conventional formalism,
we have considered four ways of constructing the cross section for each
parameterization;  considering $\Dq$ to be ${\cal O}(\alpha)$ {\em or}  
${\cal O}(\alpha/\alpha_s)$  
together with {\em either} the resummed {\em or} 
expanded solution of the evolution equation.
Provided the resummed solution of the all order evolution equation 
can be accurately determined, the approach using this solution 
and the direct terms evaluated at fixed order (approach I) represents 
the theoretically preferred approach.  
However, in Section~\ref{sec:comparison} we pointed out that the approach 
evaluating the direct terms at fixed order and using 
an expanded expression of the fragmentation function has some 
important advantages, such as eliminating the factorization scale dependence
and having an analytic form.

Predictions using these four approaches and either the GRV or BFG schemes 
were compared with
the experimental data in Section~\ref{sec:results}.
Reassuringly, in all cases, the NLO or BLL predictions 
are significantly less sensitive to the choice 
of factorization scale than the LO or LL
predictions.  We will therefore
confine our comments to comparisons of the 
NLO or BLL predictions with the data.
We note that estimates using the expanded
fragmentation function systematically lie above those using the resummed
fragmentation function.

Unfortunately, photons can be confused with neutral pions.
Consequently, the measured photon cross section can only be obtained after 
a very large experimental background subtraction has been performed.  
As a result, the experimental errors are quite large.   
In particular, the OPAL inclusive photon data is unable to discriminate
between any of the approaches or parameterizations used to
predict the cross section at NLO or BLL.
On the other hand, the ALEPH data does discriminate amongst the models,
and, apart from the very high $z$ region where the parameterization is suspect,
prefers the BFG fragmentation function.   
Estimates based on either the expanded fragmentation function 
in both GRV or BFG schemes 
or the GRV parameterization for the resummed fragmentation function 
give results that are systematically larger than
the data allows.
Nevertheless, the predictions based on the resummed
BFG parameterization, with explicit power counting 
($\Dq$ of ${\cal O}(\alpha)$) or conventional power counting
($\Dq$ of ${\cal O}(\alpha/\alpha_s)$), agree well with the data but
are too similar to be discriminated between.

To summarize,
we have shown how the inclusive and `photon' +~1 jet data from LEP can 
be described by either the fragmentation function
fitted to the ALEPH data or by the BFG solution of the evolution 
equation.
In the latter case, 
the agreement needs however to be restricted to $z$-values below 0.95. 
We expect that this good agreement can be taken across to a variety of
processes involving quarks and photons, such as
prompt photon 
production at hadron colliders and the photon pair cross section at LHC.
This may be of assistance in determining both the gluon content 
of the proton at moderate $x$ values as well as in detecting 
a Standard Model Higgs-boson of intermediate mass via its two photon decay 
at the LHC.

\section*{Acknowledgements}
We thank 
Luc Bourhis, Michel Fontannaz and Andreas Vogt 
for communications regarding the large $x$ behaviour of the
\cite{bfg} and \cite{grv} photon fragmentation functions respectively.
We also thank Thomas Gehrmann for many useful discussions.
EWNG thanks the theory groups at CERN and 
Fermilab for their kind hospitality
during the early stages of this work.
This work was supported in part by the EU Fourth Framework 
Programme `Training and Mobility of Researchers', 
Network `Quantum Chromodynamics and the Deep Structure of 
Elementary Particles', contract FMRX-CT98-0194 (DG-12 - MIHT).

\end{document}